\documentclass[12pt]{iopart}

\usepackage{iopams}  

\usepackage{graphicx}
\usepackage{pstricks}

\input epsf.sty

\begin{document}

\title{An alternative to the conventional micro-canonical ensemble}

\author{Boris V. Fine and Frank Hantschel}

\address{Institute for Theoretical Physics, 
University of Heidelberg,
Philosophenweg 19, 69120 Heidelberg, Germany}
\ead{b.fine@thphys.uni-heidelberg.de}
\begin{abstract}
Usual approach to the foundations of quantum statistical physics is based on conventional micro-canonical ensemble as a starting point for deriving Boltzmann-Gibbs (BG) equilibrium. It leaves, however, a number of conceptual and practical questions unanswered. Here we discuss these questions, thereby motivating the study of a natural alternative  known as Quantum Micro-Canonical (QMC) ensemble. We present a detailed numerical study of the properties of the QMC ensemble for finite quantum systems revealing a good agreement with the existing analytical results for large quantum systems. We also propose the way to introduce analytical corrections accounting for finite-size effects. With the above corrections, the agreement between the analytical and the numerical results becomes very accurate. The QMC ensemble leads to an unconventional kind of equilibrium, which may be realizable after strong perturbations in small isolated quantum systems having large number of levels. We demonstrate that the variance of energy fluctuations can be used to discriminate the QMC equilibrium from the BG equilibrium.   We further suggest that the reason, why BG equilibrium commonly occurs in nature rather than the QMC-type equilibrium, has something to do with the notion of quantum collapse.
\end{abstract}


\section{Introduction}

This article is about the properties of a quantum statistical ensemble, which is, in a sense, opposite to the conventional micro-canonical ensemble. It is an extension of the previous work of one of us\cite{Fine-09-statistics}. The present work contains a discussion of underlying conceptual issues and detailed analytical and numerical investigations of  finite quantum systems.  These two aspects of the article are complementary but otherwise quite different in scope and style. Sections~\ref{narrow} and \ref{prospects} are dealing with conceptual issues. They  motivate and outline a broader agenda. The remaining sections (\ref{qmc}-\ref{spins} plus Appendices) are more specialized. We hope that the conceptual part would motivate the readers unfamiliar with the above agenda to look into the rest of the article.

\section{Narrow eigenstate energy window in conventional statistical ensembles and the motivation to look beyond it}
\label{narrow}

The standard quantum-mechanical derivation of Boltzmann-Gibbs (BG) equilibrium is based on the conventional micro-canonical postulate formulated as follows. One should consider a subsystem in a macroscopic environment, assume, that the two form an isolated macroscopic system, neglect the interaction between them, and then select the statistical ensemble amounting to an incoherent mixture of eigenstates of the total system in a narrow energy window around a given value of energy, $E_{\hbox{\scriptsize av}}$. The width of this micro-canonical energy window $\Delta_{mc}$ should be large enough to include a very large number of quantum levels, but, at the same time, small enough to imply the negligible change of temperature across this energy window. This condition can be formally expressed as follows:
\begin{equation}
 \Delta_{mc} \gg {1 \over \nu(E_{\hbox{\scriptsize av}})},
\label{Dnu}
\end{equation}
and
\begin{equation}
 {1 \over T} \equiv 
\left. { \partial \hbox{log} [\nu(E)/\nu_0] \over \partial E }
\right\vert_{E = E_{\hbox{\scriptsize av}}}
\gg
- \Delta_{mc} \left. { \partial^2 \hbox{log} [\nu(E)/\nu_0] \over \partial E^2 }
\right\vert_{E = E_{\hbox{\scriptsize av}}}.
\label{T}
\end{equation}
Here $\nu(E)$ is the density of energy states of the isolated macroscopic system, which should be approximately equal to the density of states of the environment at the same energy, $\nu_0$ is unimportant normalization constant, and $T$ is temperature in energy units. Given the exponentially large number of levels of a macroscopic system, the above two inequalities can be easily satisfied for $\Delta_{mc}$ ranging over many orders of magnitude.

Recently, further progress was made by showing that a single pure state of a macroscopic system formed by a coherent superposition of eigenstates randomly selected from the above energy window would also lead to a density matrix of the subsystem exhibiting the BG distribution\cite{Gemmer-04,Popescu-06,Goldstein-06,Reimann-08}. It has been further conjectured that, if the isolated macroscopic system is chaotic in the classical limit, then even a single eigenstate would lead for the BG distribution in the density matrix of the subsystem\cite{Srednicki-94}.

Yet the question remains: What is the justification of the narrow energy window in the microcanonical postulate? The origin, the limits and the implications of this condition are not well understood at present.  

The microcanonical postulate as such remains a rather formal construction, because in reality we do not encounter macroscopic systems isolated from the rest of the world. Even, if we were to find one, the non-relativistic quantum-mechanical description of such a system normally leads to unrealistically small spacings between adjacent eigenstates --- much smaller than the minimal natural width of an energy level estimated as  $\hbar / t_U$, where $t_U \approx 4 \times 10^{17}$s is the Big-Bang age of the Universe. The real width of the energy levels should, of course, be much larger due to relativistic effects. Although the notion of an energy eigenstate of a macroscopic system is not well-defined, one can still hope that the statistical treatment based on this notion would produce an outcome, whose validity extends beyond the systems with well-defined energy eigenstates --- in the same way as BG statistics derivable for classical systems extends to quantum systems. 

Considering systems with well-defined eigenstates, one can proceed by giving the conventional microcanonical ensemble with its narrow energy window the status of a fundamental postulate. However, such an approach to justifying BG equilibrium is not quite satisfactory.  If instead of simply postulating the BG equilibrium for quantum systems, one were to claim its derivation from a more fundamental postulate, then the introduction of this extra postulate would be justified only if, in addition to the BG equilibrium, it would lead to an explanation of some other previously unexplained facts about nature and predict new ones. The micro-canonical postulate, however, does not lead to any other observable prediction besides BG equilibrium itself. 

It could have been that the narrow energy condition is only a simplification, while the BG distribution emerges for a broader class of ensembles. However, as discussed in Refs.\cite{Aarts-00,Fine-09-statistics,Deutsch-09}, the departure from the narrow energy condition easily spoils the BG statistics for a subsystem. It leads to a mixture of different thermal distributions, which does not produce a distribution characterized by a single temperature.  Yet one could have asked the question: If it is not a narrow energy window, then what is the most likely shape of the participation function for the energy eigenstates of the entire system? Can it be that, if the most likely shape is found, then the resulting statistics for the subsystems is of BG type?

One can, in particular, recall that the conventional canonical ensemble does not appear to have narrow energy window for participating eigenstates. To each eigenstate with energy $E$, it assigns probability weight  $p(E) \cong \hbox{exp}(-E/T_0)$, which follows either from maximizing entropy $-\sum_i p(E_i) \hbox{log} \ p(E_i)$ subject to the total energy constraint, or from the practical consideration that, if one slowly isolates a large quantum system from its environment, which initially was in BG equilibrium with temperature $T_0$, then the diagonal elements of the resulting density matrix will have the above form. Here, it should be noted, however, that the quantity of practical interest in not the probability weight function  $p(E)$ as such  but rather its product with the density of states $p(E) \nu(E)$. Since $p(E)$ decays exponentially with constant ``rate" $1/T_0$, while  for a macroscopic system $\nu(E)$ increases exponentially with decreasing "rate" $1/T(E)$, where $T(E)$ is defined by Eq.(\ref{T}), the product would have a narrow maximum at energy $E = E_{\hbox{\scriptsize av}}$ corresponding to $T(E_{\hbox{\scriptsize av}}) = T_0$. The width of that maximum $\Delta_c$ is determined by
\begin{equation}
\left. { \partial^2 \hbox{log} [\nu(E)/\nu_0] \over \partial E^2 }
\right\vert_{E = E_{\hbox{\scriptsize av}}} = 
\left. { \partial [1/T(E)] \over \partial E }
\right\vert_{E = E_{\hbox{\scriptsize av}}}
 \sim - {1 \over T_0^2 N_p},
\label{plog}
\end{equation}
where $N_p$ is the number of particles in the system. This gives 
 $\Delta_c \sim T_0 \sqrt{N_p}$. As a result, 
\begin{equation}
 - \Delta_c \left. { \partial^2 \hbox{log} [\nu(E)/\nu_0] \over \partial E^2 }
\right\vert_{E = E_{\hbox{\scriptsize av}}} \sim {1 \over T_0 \sqrt{N_p}} \ll {1 \over T_0},
\label{Dlog}
\end{equation}
i.e. condition (\ref{T}) is satisfied. Thus  the canonical ensemble for an isolated macroscopic quantum system, in effect, imposes a narrow energy window condition.

The idea of maximization entropy at a given energy is closely related to the notion of chaos and ergodicity in the phase space of classical systems. For quantum systems, the notion of phase space trajectories is not defined because of the Heisenberg uncertainty principle. Hence the maximization of conventional entropy cannot be justified on the basis of the dynamics of isolated quantum systems.

To address the issue of what may happen beyond the narrow energy window imposed by the conventional ensembles, an ensemble with unrestricted participation of eigenstates and without the requirement of the maximization of conventional entropy, referred to below as Quantum Micro-Canonical ensemble\cite{Brody-98,Brody-05,Bender-05,Fine-09-statistics}, is a conceptually necessary limit to consider. (Similar quantum canonical ensembles have also been defined Refs.\cite{Brody-98,Jona-Lasinio-06}.)

In addition to the above conceptual issues, the basic concern in the context of the foundations of quantum statistical physics remains about the actual behavior of an isolated quantum system perturbed from equilibrium and not interacting with a large thermal bath. 
It is overwhelming intuition of many researchers, including the present authors, that an isolated macroscopic classical system with generic non-linear interaction between particles would evolve dynamically to exhibit the BG statistics for small subsystems.
The situation is, however, not so trivial for quantum systems, because they can be in a superposition of states having different total energies. If, after the perturbation, an isolated quantum system is in a broad superposition of eigenstates violating the narrow energy condition, then the occupations numbers of eigenstates will not change with time under the dynamics guided by linear Schr\"{o}dinger equation, and hence the narrow energy condition  will never emerge.

\section{Quantum micro-canonical ensemble}
\label{qmc}

Generic wave function $\Psi$ of a quantum system is a superposition of eigenstates $\phi_i$:
\begin{equation}
\Psi = \sum_{i=1}^{N} C_i \phi_i,
\label{Psi}
\end{equation}
where $C_i$ are complex amplitudes, and $N$ the total number of eigenstates.
For an isolated quantum system, fixing energy $E_{\hbox{\scriptsize av}}$ associated with wave function $\Psi$ implies 
\begin{equation}
\sum_{i=1}^{N}  E_i p_i  = E_{\hbox{\scriptsize av}},
\label{epsav}
\end{equation}
where $p_i = |C_i|^2$ are the occupation numbers of quantum eigenstates, and $E_i$ are the corresponding energies. We shall refer to $E_{\hbox{\scriptsize av}}$ as ``average energy'' [not to be confused with the uniform average of all $E_i$].

In this work, we focus on Quantum Micro-Canonical (QMC) ensemble\cite{Brody-98,Brody-05,Bender-05,Fine-09-statistics}, where all quantum wave functions satisfying average energy constraint (\ref{epsav}) are equally probable in the Hilbert space of the system. The QMC ensemble amounts to a natural generalization of the equipartition postulate for energy shells in classical phase spaces. Despite this analogy, the QMC ensemble is quite different from the conventional micro-canonical ensemble, because it does not limit the participating eigenstates to a narrow energy window around $E_{\hbox{\scriptsize av}}$.

It was shown in Ref.\cite{Fine-09-statistics} that the QMC ensemble does not lead to the BG statistics, at least in the absence of interaction between the subsystem and the environment. This negative result is important, because it forces one to think seriously about justifying the narrow energy window in the conventional microcanonical ensemble. As discussed in Refs.\cite{Fine-09-statistics,Deutsch-09} and in Section~\ref{prospects}, quantum collapse may play here a fundamental role.  Further study of the QMC ensemble is also important, because this ensemble may be realizable in small isolated quantum systems with a large number of quantum levels.  

Beyond the subject of foundations of quantum statistical physics, the properties of the QMC ensemble are also relevant to the studies of various properties of Hilbert spaces, e.g. in the context of quantum computing or many-particle entanglement.

It is finally plausible on general grounds that such a natural quantum ensemble would find many realizations not discovered so far. For example, it was demonstrated in Ref.\cite{Fine-09-statistics} that the expansion of the basis states of sparce random matrices in terms of eigenstates exhibits the QMC type of statistics.

The description of the QMC ensemble involves two steps: (1) characterization of the shape of the energy window function --- in this case, the participation function for the eigenstates of the entire isolated system; and (2) obtaining the density matrix for the subsystem of interest. Sections~\ref{general}-\ref{results} below deal with step (1), while Section~\ref{spins} deals with step (2).

In Section~\ref{general}, we summarize the analytical results obtained in Ref.\cite{Fine-09-statistics} for the QMC ensemble in large-$N$ isolated quantum systems. In Section~\ref{ansatz}, we propose a method of introducing finite-$N$ corrections to the large-$N$ approximation. In Section~\ref{results}, we present the numerical tests of the resulting description for a 12-level isolated quantum system. In Section~\ref{spins}, we examine numerically the density matrix for a subsystem of a 12-level system, and compare the QMC-based result with that of the conventional canonical ensemble.  Section~\ref{prospects} discusses open questions motivated by our results.   \ref{algorithmsL} describes a practical numerical algorithm for finding an important characteristic function appearing in the course of the analytical approximation procedure. \ref{algorithmMC} gives the details of our Monte-Carlo sampling algorithm for the QMC ensemble.

\section{Description of the previous QMC results}
\label{general}

We consider quantum spectum $\{ E_i \}$ consisting of $N$ energy levels. The values of energies are assumed to be ordered with $E{\hbox{\scriptsize min}} \equiv E_1$ and $E{\hbox{\scriptsize max}} \equiv E_N $.

The analytical approach and the analytical results of Ref.\cite{Fine-09-statistics} can be described as follows.

The QMC ensemble is defined by specifying uniform joint probability distribution for complex amplitudes $C_i \equiv |C_i| e^{i \varphi_i}$ in the Hilbert space --- subject to average energy constraint (\ref{epsav}) and to the normalization constraint
\begin{equation}
\sum_i^N p_i = 1.
\label{norm}
\end{equation}
The two constraints are of the second order in variables $|C_i|$ and do not depend on $\varphi_i$. 

To avoid dealing with curved manifolds, we now change from variables $|C_i|$ to variables $p_i = |C_i|^2$. Completely random choice of states in Hilbert space implies that the probability measure is proportional to the volume element:
\begin{equation}
dV = \prod_i^N |C_i| \  d \varphi_i \  d |C_i| = 
{1 \over 2} \prod_i^N  d \varphi_i \  d (|C_i|^2) = {1 \over 2} \prod_i^N  d \varphi_i \  d p_i.
\label{dV0}
\end{equation}
Since constraints (\ref{epsav}, \ref{norm}) do not depend on $\varphi_i$, the assignment of variables $\varphi_i$ is totally random. The resulting uniform distribution of $\varphi_i$ can be just factorized out and then integrated. What remains is the Euclidean subspace of variables $p_i$ with uniform joint probability measure on the manifold constrained by conditions (\ref{epsav}, \ref{norm}) together with the positivity condition
\begin{equation}
p_i \geq 0,\ \ \ \ \  \forall i.
\label{positiv}
\end{equation}
In the $p_i$-space, due to the linear nature of constraints (\ref{epsav},\ref{norm},\ref{positiv}), the boundaries of the resulting manifold are flat, i.e. the manifold itself is an $(N-2)$-dimensional polyhedron. We will refer to it as ``QMC polyhedron''.  Our goal is to obtain marginal probability distribution $P_k(p_k)$ for a given occupation number $p_k$ and the corresponding average value $\langle p_k \rangle$ representing the participation of the $k$th eigenstate in the QMC ensemble.   

Probability distribution $P_k(p_k)$ is proportional to the $(N-3)$ dimensional volume of the intersection of the QMC polyhedron with hyperplane $p_k = const$. (Here we use variable $p_k$ as a label of the coordinate axis, while in $P_k(p_k)$ and everywhere below, $p_k$ refers to the value of $const$.) The intersection manifold is to be denoted as ${\cal M}_k$, and its volume as $V_k(p_k)$. This volume plays the role of unnormalized probability distribution such that
\begin{equation}
P_k(p_k) = { V_k(p_k) \over \int_0^1 V_k(p_k^{\prime})dp_k^{\prime}},
\label{Pkpk}
\end{equation}
and the average value of $p_k$ is 
\begin{equation}
\langle p_k \rangle = {
\int_0^1 p_k^{\prime} V_k(p_k^{\prime})dp_k^{\prime} 
\over 
\int_0^1 V_k(p_k^{\prime})dp_k^{\prime} }.
\label{pkav}
\end{equation}

In the space of all variables excluding $p_k$ --- to be denoted as $\{ p_i \}_k$, manifold ${\cal M}_k$ is defined by conditions:
\begin{equation}
\sum_{i, i\neq k}^N p_i = 1 - p_k,
\label{normK}
\end{equation}
and
\begin{equation}
\sum_{i, i\neq k}^{N}  (E_i - E_{\hbox{\scriptsize av}})  p_i   =  - (E_k - E_{\hbox{\scriptsize av}}) p_k,
\label{epsavK0}
\end{equation}
together with positivity conditions (\ref{positiv}).

The most relevant limit to consider now is $p_k \ll 1$, because, for $N \gg 1$, it is improbable that a single occupation number $p_k$ in a randomly chosen superposition of all quantum states becomes comparable to 1. We refer to this limit as ``small-$p_k$ approximation.'' It gives\cite{Fine-09-statistics}
\begin{equation}
V_k(p_k) = V_k(0) e^{-N p_k[1 + \lambda_k (E_k - E_{\hbox{\scriptsize av}})]}.
\label{Vkpk}
\end{equation}
where $\lambda_k$ is the volume renormalization parameter associated with the shift of energy hyperplane (\ref{epsavK0}).

It was further shown in Ref.\cite{Fine-09-statistics} that, in the limit $N \gg 1$ all parameters $\lambda_k$ are approximately equal to the same value --- to be denoted as $\lambda$, i.e. in this limit
\begin{equation}
V_k(p_k) = V_k(0) e^{-N p_k[1 + \lambda (E_k - E_{\hbox{\scriptsize av}})]}.
\label{Vkpk1}
\end{equation}
The corresponding average value of the occupation number $p_k$ is
\begin{equation}
\langle p_k \rangle = {1 \over N [1 + \lambda (E_k - E_{\hbox{\scriptsize av}})]}.
\label{pav2}
\end{equation}
The value of parameter $\lambda$ can now be found numerically by substituting Eq.(\ref{pav2}) into either averaged normalization condition
\begin{equation}
\sum_{k=1}^N \langle p_k \rangle = 1,
\label{normav}
\end{equation}
or averaged energy condition
\begin{equation}
\sum_{k=1}^N  (E_k - E_{\hbox{\scriptsize av}}) \langle p_k \rangle   =  0.
\label{epsavav}
\end{equation}
The two values of $\lambda$ are guaranteed to be the same for any $N$, probably because formula (\ref{pav2}) is analytically accurate leading term in the large-$N$ expansion. [For finite-$N$ systems, the actual behavior of $V_k(p_k)$ can exhibit noticeable deviations from the exponential dependence of Eq.(\ref{Vkpk1})].

Behind the geometry-based discussion one should not overlook that Eq.(\ref{Vkpk1}) represents grand-canonical Gibbs ensemble defined in the Hilbert space, or the space of occupation numbers $p_k$. The problem is, in fact, equivalent to finding the probability distributions for a system of $N$ classical states characterized by one-particle energies $E_k$, with $p_k$ being the number of classical particles occupying the $k$-th state. In such a problem both the total energy and the total number of classical particles is conserved, which leads to the grand canonical ensemble in the Hilbert space. This analogy, however, does not imply conventional observable results, because it pertains to the probability distribution $P_k(p_k)$, while, in the original quantum problem, $p_k$ itself is the probability of observing the system in the $k$th state, i.e. the prediction of the observable outcome requires finding the average value $\langle p_k \rangle$. It is at this last step that the departure from the conventional statistics occurs. In a broader context, this kind of result associated with double-averaging is known as superstatistics\cite{Beck-03}.

The grand-canonical analogy becomes evident, once formula (\ref{Vkpk1}) is rewritten as 
\begin{equation}
V_k(p_k) = V_k(0) \hbox{exp} \{-N \lambda (E_k p_k - E_{\lambda} p_k) \},
\label{Vkpk_grand}
\end{equation}
where 
\begin{equation}
 E_{\lambda} = E_{\hbox{\scriptsize av}} - {1 \over \lambda}.
\label{elambda}
\end{equation}
Equation (\ref{Vkpk_grand}) implies that $\lambda N$ is the inverse temperature in the Hilbert space, and $E_{\lambda}$ is the chemical potential.

The value of $\lambda$ is zero, when $E_{\hbox{\scriptsize av}}$ is equal to $E_{\hbox{\scriptsize av}0}$, the uniform average of all energies in spectrum $\{ E_i \}$\cite{Bender-05,Fine-09-statistics}:
\begin{equation}
 E_{\hbox{\scriptsize av}0} = {1\over N} \sum_{i=1}^N E_i.
\label{Eav0}
\end{equation}
We also recall that $\lambda > 0$ for $E_{\hbox{\scriptsize av}} < E_{\hbox{\scriptsize av}0} $, and
$\lambda < 0$ for $E_{\hbox{\scriptsize av}} > E_{\hbox{\scriptsize av}0} $.
Here we follow the convention of Ref.\cite{Fine-09-statistics}, that, unless specified otherwise, the origin of the energy axis is set at $E_{\hbox{\scriptsize av}0}$, i.e., $E_{\hbox{\scriptsize av}0}=0$.

The chemical potential $E_{\lambda}$ is simultaneously the pole of function $\langle p_k \rangle (E_{\hbox{\scriptsize av}})$ given by Eq.(\ref{pav2}), which implies that for the small-$p_k$ approximation to be valid for all levels, $E_{\lambda}$ should be sufficiently below the lowest energy level for positive $\lambda$  or above the highest energy level for negative $\lambda$. In the former case, ``sufficiently below''  means that\cite{Fine-09-statistics} 
\begin{equation}
 E_{\hbox{\scriptsize min}} - E_{\lambda} \gg {E_{\hbox{\scriptsize av}} - E_{\hbox{\scriptsize min}}\over N}.
\label{ElEmin}
\end{equation}
It is shown in Ref.\cite{Fine-09-statistics} that the above condition is violated for a typical macroscopic system at a typical value of $E_{\hbox{\scriptsize av}}$, which leads to condensation into the lowest energy state, i.e. $\langle p_1 \rangle$ becomes comparable to one. Condition (\ref{ElEmin}) can also be easily violated for finite-N systems leading to the departure from the small-$p_k$ results (\ref{Vkpk1},\ref{pav2}) for several lowest levels. In the both cases, it is necessary to consider what happens beyond the small-$p_k$ approximation.

When typical values of $p_k$ are not small, it is necessary to recall \cite{Fine-09-statistics} that parameters $\lambda_k$ are defined as functions of average energy $E_{\hbox{\scriptsize av}}$ as follows: 
\begin{equation}
 \lambda_k[E_{\hbox{\scriptsize av}}] = {1 \over V_k} 
                     \left. {\partial V_k \over \partial v} \right|_{v=0},
\label{lk}
\end{equation}
where [in this equation only] $V_k$ is the volume of the manifold constrained in space $\{p_i\}_k$ by conditions
\begin{equation}
\sum_{i, i\neq k}^N p_i = 1 ,
\label{normK1}
\end{equation}
\begin{equation}
\sum_{i, i\neq k}^{N}  (E_i - E_{\hbox{\scriptsize av}})  p_i   =  v,
\label{epsavK01}
\end{equation}
together with positivity conditions (\ref{positiv}). Here $v$ is the energy shift parameter. We follow the convention of Ref.\cite{Fine-09-statistics} to use square brackets in functions $\lambda_k[E_{\hbox{\scriptsize av}}]$  and later in $\lambda[E_{\hbox{\scriptsize av}}]$.

Given the above definition of $\lambda_k[E_{\hbox{\scriptsize av}}]$, the volume of manifold ${\cal M}_k$ defined by Eqs.(\ref{positiv},\ref{normK},\ref{epsavK0}) can be expressed as
\begin{equation}
 V_k(p_k) = V_k(0) \hbox{exp} \left\{
(N-3)
\left[ \hbox{log}(1-p_k) + \int_{E_{\hbox{\scriptsize av}}}^{E_{\hbox{\scriptsize av}} - {(E_k -E_{\hbox{\scriptsize av}}) p_k \over 1-p_k} } \lambda_k[E] d E
\right]
\right\}.
\label{Vkpk2A}
\end{equation}
[This equation does not appear explicitly in Ref.\cite{Fine-09-statistics}, but otherwise it is an obvious intermediate step in the calculation described there.]
In the limit $N \gg 1$, all $\lambda_k[E]$ are approximately equal to a single function $\lambda[E]$ defined in interval $[E_{\hbox{\scriptsize min}}, E_{\hbox{\scriptsize max}}]$. Therefore, Eq.(\ref{Vkpk2A}) becomes\cite{Fine-09-statistics}
\begin{equation}
 V_k(p_k) = V_k(0) \hbox{exp} \left\{
(N-3)
\left[ \hbox{log}(1-p_k) + \int_{E_{\hbox{\scriptsize av}}}^{E_{\hbox{\scriptsize av}} - {(E_k -E_{\hbox{\scriptsize av}}) p_k \over 1-p_k} } \lambda[E] d E
\right]
\right\}.
\label{Vkpk2}
\end{equation}
Functions $V_k(p_k)$ given by both Eq.(\ref{Vkpk2A}) and (\ref{Vkpk2}) are not necessarily defined in the entire range $0 \leq p_k \leq 1$, because, for too large $p_k$, and, in the case of $k=1$ or $k=N$, for too small $p_k$, the intersection manifold defined by Eqs.(\ref{positiv},\ref{normK},\ref{epsavK0}) may simply not exist.  The practical rule is that $V_k(p_k)$ is defined, when $\lambda_k[E]$ or $\lambda[E]$ are defined at the upper integration limit  $E_{\hbox{\scriptsize av}} - {(E_k -E_{\hbox{\scriptsize av}}) p_k \over 1-p_k}$. This limit should fall within interval $[E_{\hbox{\scriptsize min}}, E_{\hbox{\scriptsize max}}]$ for Eq.(\ref{Vkpk2}) and within sometimes different boundaries (given in Section~\ref{ansatz}) for Eq.(\ref{Vkpk2A}). The upper cutoffs for $p_k$ are given in Appendix~D of Ref.\cite{Fine-09-statistics} and the lower cutoffs in Section~\ref{ansatz}. The mean-$\lambda$ approximation represented by Eq.(\ref{Vkpk2}) neglects the existence of lower cutoffs of $p_k$ for $k=1$ and $k=N$. These cutoffs, however, can be incorporated, at the level of finite-$N$ corrections based on Eq.(\ref{Vkpk2A}).  If a lower cutoff is present, then $V_k(0)$ is not defined, and the prefactor in Eq.(\ref{Vkpk2A}) should be viewed as simply a constant, which is then canceled by normalization in Eq.(\ref{Pkpk}). Also, when the lower cutoff is present, the lower integration limit ``$E_{\hbox{\scriptsize av}}$'' in Eq.(\ref{Vkpk2A}) should be replaced with the minimum or the maximum value of energy, where $\lambda_k[E]$ is defined, for $k=1$ and $k=N$, respectively. In general, function $V_k(p_k)$ becomes exponentially small as it approaches either of its cutoffs.  Therefore, as $N$ increases, the practical significance of these cutoffs decreases.

In Eq.(\ref{Vkpk2}), we took, at first sight inconsistently, the large-$N$ limit $\lambda_k[E]\approx \lambda[E]$ but did not approximate $N-3 \approx N$. The reason is that we are aiming at making numerical tests for Hilbert space with $N \sim 10$, where condition $N-3 \gg 1$ is not well fulfilled, and, since the term $(N-3) \hbox{log}(1-p_k)$ is an exact one\cite{Fine-09-statistics}, we thereby limit the finite-$N$ error only to the integral in Eq.(\ref{Vkpk2}) and deal with this error later in Section~\ref{ansatz}. It is also worth noting that the small-$p_k$ formulas (\ref{Vkpk}) and (\ref{Vkpk1}) can be obtained by linearizing exponents in 
Eqs.(\ref{Vkpk2A}) and (\ref{Vkpk2}) with respect to $p_k$, and approximating $N-3 \approx N$.

Equation (\ref{Vkpk2}) was the basis for proving the condensation of the QMC ensemble into the lowest energy state of macroscopic systems\cite{Fine-09-statistics}. It indicated that the character of function $V_1(p_1)$ was not exponential but rather narrowly peaked around $p_1 = \langle p_1 \rangle$. This condensation is different from the Bose-Einstein condensation, in the sense that it occurs in the many-particle ground state independently of the type of the particles composing the system.

\section{Ansatz for deviations of $\lambda_k[E_{\hbox{\scriptsize av}}]$ from $\lambda[E_{\hbox{\scriptsize av}}]$}
\label{ansatz}

While we cannot propose a controllable procedure for finding corrections for $\lambda_k[E_{\hbox{\scriptsize av}}]$ with respect to $\lambda[E_{\hbox{\scriptsize av}}]$, we introduce here an ansatz, which makes these corrections in a conceptually meaningful way and, at the same time, significantly improves the agreement with Monte-Carlo results.

The ansatz is based on the following considerations. Function $\lambda[E_{\hbox{\scriptsize av}}]$ is defined in interval $[E_{\hbox{\scriptsize min}}, E_{\hbox{\scriptsize max}}]$. It has two universal properties in the large-$N$ limit\cite{Fine-09-statistics}:

(i) $\lambda[E_{\hbox{\scriptsize av}0}] = 0$;

(ii) for $E_{\hbox{\scriptsize av}} < E_{\hbox{\scriptsize av}0}$, $\lambda[E_{\hbox{\scriptsize av}}]$ quickly approaches the asymptotic form:
\begin{equation}
\lambda[E_{\hbox{\scriptsize av}}] \approx {1 \over E_{\hbox{\scriptsize av}} - E_{\hbox{\scriptsize min}} },
\label{lambdaA}
\end{equation}
while for $E_{\hbox{\scriptsize av}} > E_{\hbox{\scriptsize av}0}$, the asymptotic form is
\begin{equation}
\lambda[E_{\hbox{\scriptsize av}}] \approx {1 \over E_{\hbox{\scriptsize av}} - E_{\hbox{\scriptsize max}} }.
\label{lambdaB}
\end{equation}

In other words, the behavior of the entire function $\lambda[E_{\hbox{\scriptsize av}}]$ is almost entirely determined just by the values of three parameters $E_{\hbox{\scriptsize av}0}$, $E_{\hbox{\scriptsize min}}$ and $E_{\hbox{\scriptsize max}}$.

Now, function $\lambda[E_{\hbox{\scriptsize av}}]$ characterizes the entire quantum spectrum $\{E_i\}$,
while functions $\lambda_k[E_{\hbox{\scriptsize av}}]$ characterize spectra $\{E_i\}_k$ obtained from $\{E_i\}$ by removing energy level $E_k$. As a result, the uniform average of all energies for spectrum $\{E_i\}_k$  is shifted to 
\begin{equation}
 E_{\hbox{\scriptsize av}0k} = E_{\hbox{\scriptsize av}0} + 
{ E_{\hbox{\scriptsize av}0} - E_k \over N-1},
\label{Eavok}
\end{equation}
while the end points of the spectrum remain unchanged, {\it unless} $k=1$ or $k=N$, when, respectively, the minimum or the maximum energy levels are removed:
\begin{equation}
 E_{\hbox{\scriptsize min}k} = \left\{
\begin{array}{l@{\quad:\quad}l}
E_2 & k=1 
\\ 
E_{\hbox{\scriptsize min}} & k > 1 ;
\end{array}
\right. 
\label{Emink}
\end{equation}
\begin{equation}
 E_{\hbox{\scriptsize max}k} = \left\{
\begin{array}{l@{\quad:\quad}l}
E_{\hbox{\scriptsize max}} & k < N, 
\\ 
E_{N-1} & k=N .
\end{array}
\right. 
\label{Emaxk}
\end{equation}
 
Apart from the above specific differences $\lambda_k[E_{\hbox{\scriptsize av}}]$ should be reminiscent of $\lambda[E_{\hbox{\scriptsize av}}]$. Therefore, in order to approximate $\lambda_k[E_{\hbox{\scriptsize av}}]$ from the knowledge of $\lambda[E_{\hbox{\scriptsize av}}]$ we adopt the following two-step procedure:

(I) Shift of $\lambda[E_{\hbox{\scriptsize av}}]$ to move $E_{\hbox{\scriptsize av}0}$ to $E_{\hbox{\scriptsize av}0k}$:
\begin{equation}
 \lambda_k^{\prime}[E_{\hbox{\scriptsize av}}] = \lambda[E_{\hbox{\scriptsize av}} + E_{\hbox{\scriptsize av}0} - E_{\hbox{\scriptsize av}0k}].
\label{Lshift}
\end{equation}
This also shifts $E_{\hbox{\scriptsize min}}$ and $E_{\hbox{\scriptsize max}}$ (the end points of $\lambda[E_{\hbox{\scriptsize av}}]$) to 
$E_{\hbox{\scriptsize min}k}^{\prime} = E_{\hbox{\scriptsize min}} + E_{\hbox{\scriptsize av}0} - E_{\hbox{\scriptsize av}0k} $ and 
$E_{\hbox{\scriptsize max}k}^{\prime} = E_{\hbox{\scriptsize max}} + E_{\hbox{\scriptsize av}0} - E_{\hbox{\scriptsize av}0k} $, respectively.

(II) Uniform contraction or expansion of energy differences  above and below $E_{\hbox{\scriptsize av}0k}$  to shift $E_{\hbox{\scriptsize min}k}^{\prime}$ and $E_{\hbox{\scriptsize max}k}^{\prime}$  to, respectively $E_{\hbox{\scriptsize min}k}$ and $E_{\hbox{\scriptsize max}k}$ given by Eqs.(\ref{Emink},\ref{Emaxk}) and simultaneously rescale $\lambda_k^{\prime}$ itself to make sure that,  if $\lambda_k^{\prime}[E_{\hbox{\scriptsize av}}]$ has asymptotic forms
${1 \over E_{\hbox{\scriptsize av}} - E_{\hbox{\scriptsize min}k}^{\prime} }$ above $E_{\hbox{\scriptsize min}k}^{\prime}$ and
${1 \over E_{\hbox{\scriptsize av}} - E_{\hbox{\scriptsize max}k}^{\prime} }$ below
$E_{\hbox{\scriptsize max}k}^{\prime}$, then the resulting function $\lambda_k[E_{\hbox{\scriptsize av}}]$ has asymptotic forms ${1 \over E_{\hbox{\scriptsize av}} - E_{\hbox{\scriptsize min}k} }$ above $E_{\hbox{\scriptsize min}k}$ and
${1 \over E_{\hbox{\scriptsize av}} - E_{\hbox{\scriptsize max}k} }$ below
$E_{\hbox{\scriptsize max}k}$. This transformation has form
\begin{equation}
 \lambda_k[E_{\hbox{\scriptsize av}}] = \left\{
\begin{array}{l@{\quad:\quad}l}
{1 \over \alpha_-} 
\lambda_k^{\prime}[E_{\hbox{\scriptsize av}0k} + \alpha_- (E_{\hbox{\scriptsize av}} - E_{\hbox{\scriptsize av}0k})]
& 
E_{\hbox{\scriptsize av}} < E_{\hbox{\scriptsize av}0k} ,
\\ 
&
\\
{1 \over \alpha_+} 
\lambda_k^{\prime}[E_{\hbox{\scriptsize av}0k} + \alpha_+ (E_{\hbox{\scriptsize av}} - E_{\hbox{\scriptsize av}0k})]
& 
E_{\hbox{\scriptsize av}} > E_{\hbox{\scriptsize av}0k} ,
\end{array}
\right. 
\label{lambdak}
\end{equation}
where
\begin{equation}
 \alpha_- = {E_{\hbox{\scriptsize min}k} - E_{\hbox{\scriptsize av}0k} \over E_{\hbox{\scriptsize min}k}^{\prime} - E_{\hbox{\scriptsize av}0k} },
\label{alpha-}
\end{equation}
and 
\begin{equation}
 \alpha_+ = {E_{\hbox{\scriptsize max}k} - E_{\hbox{\scriptsize av}0k} \over E_{\hbox{\scriptsize max}k}^{\prime} - E_{\hbox{\scriptsize av}0k} }.
\label{alpha+}
\end{equation}
The resulting function $\lambda_k[E_{\hbox{\scriptsize av}}]$ is then used in Eq.(\ref{Vkpk2A}).

The last step automatically corrects the drawback of the mean-$\lambda$ formula (\ref{Vkpk2}) mentioned in Section~\ref{general}. Namely, the mean-$\lambda$ approximation does not have lower cutoffs for $V_1(p_1)$ and $V_N(p_N)$, while the exact solution should have them.   Indeed, if $E_{\hbox{\scriptsize av}}$ falls between $E_{\hbox{\scriptsize min}}$ and $E_2$, then $V_1(p_1)$ has the lower cuttoff
\begin{equation}
 p_{1\hbox{\scriptsize min}} = {E_2 - E_{\hbox{\scriptsize av}} \over E_2 - E_{\hbox{\scriptsize min}}},
\label{p1min}
\end{equation}
which corresponds to only two lowest levels $E_{\hbox{\scriptsize min}}$ and 
$E_2$ being occupied with the occupation numbers determined by the value of $E_{\hbox{\scriptsize av}}$. Once higher energy levels become occupied, the value of $p_1$ becomes greater than $p_{1\hbox{\scriptsize min}}$ to fulfill average energy condition (\ref{epsav}). Likewise, if $E_{\hbox{\scriptsize av}}$ falls  between $E_{N-1}$ and $E_{\hbox{\scriptsize max}}$, then lower cutoff for $V_N(p_N)$ is
\begin{equation}
 p_{N\hbox{\scriptsize min}} = {E_{N-1} - E_{\hbox{\scriptsize av}} \over E_{N-1} - E_{\hbox{\scriptsize max}}}.
\label{pNmin}
\end{equation}
In our approximation procedure, $V_k(p_k)$ is defined only, when $\lambda_k [E]$ is defined at the upper integration limit in Eq.(\ref{Vkpk2A}), i.e. when
$E_{\hbox{\scriptsize av}} - {(E_k -E_{\hbox{\scriptsize av}}) p_k \over 1-p_k}$ falls in the interval between 
$E_{\hbox{\scriptsize min}k}$ and $E_{\hbox{\scriptsize max}k}$. 
The appearance of $E_{\hbox{\scriptsize min}1} \neq E_{\hbox{\scriptsize min}} $ and $E_{\hbox{\scriptsize max}N} \neq E_{\hbox{\scriptsize max}} $ in Eqs.(\ref{alpha-},\ref{alpha+}) then implies the correct lower cuttoff values $p_{1\hbox{\scriptsize min}}$ and $p_{N\hbox{\scriptsize min}}$, respectively. 
These lower cutoffs are unimportant in the large-$N$ limit, but become important for finite-$N$ systems, like the one considered numerically in this work.

\section{Numerical investigation of the QMC ensemble for a finite isolated quantum system}
\label{results}

Here we compare analytical and numerical results for a quantum spectrum that has $N=12$ energy levels. This is nearly the maximum number of quantum levels, whose statistics we can access in reasonable time with the Monte-Carlo algorithm described in \ref{algorithmMC}. The values of the spectrum energies are \{$-1.27$, $-0.93$, $-0.68$, $-0.47$, $-0.27$, $-0.09$, $0.09$, $0.27$, $0.47$, $0.68$, $0.93$, $1.27$\} (also shown in Fig.~1).  This spectrum is obtained as a strongly discretized version of the Gaussian density of states with maximum at $E_{\hbox{\scriptsize av}0} = 0$. (Nearly Gaussian density of states is generically expected for macroscopic systems\cite{Fine-09-statistics}.)

\begin{figure}[t] \setlength{\unitlength}{1.0mm}
 
\psset{xunit=0.1cm,yunit=0.1cm}
\begin{pspicture}(0,0)(157,80)
{
\put(0,0){ \epsfxsize= 4.5in \epsfbox{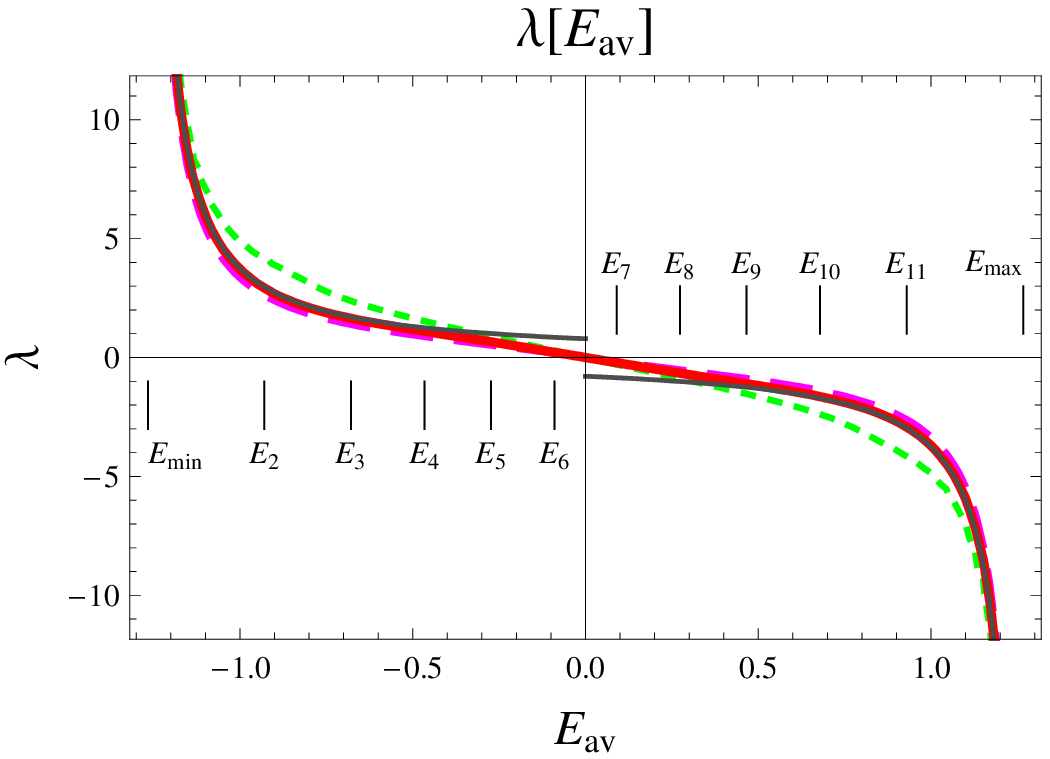} }
\put(2.4,5){ \epsfxsize= 1.55in \epsfbox{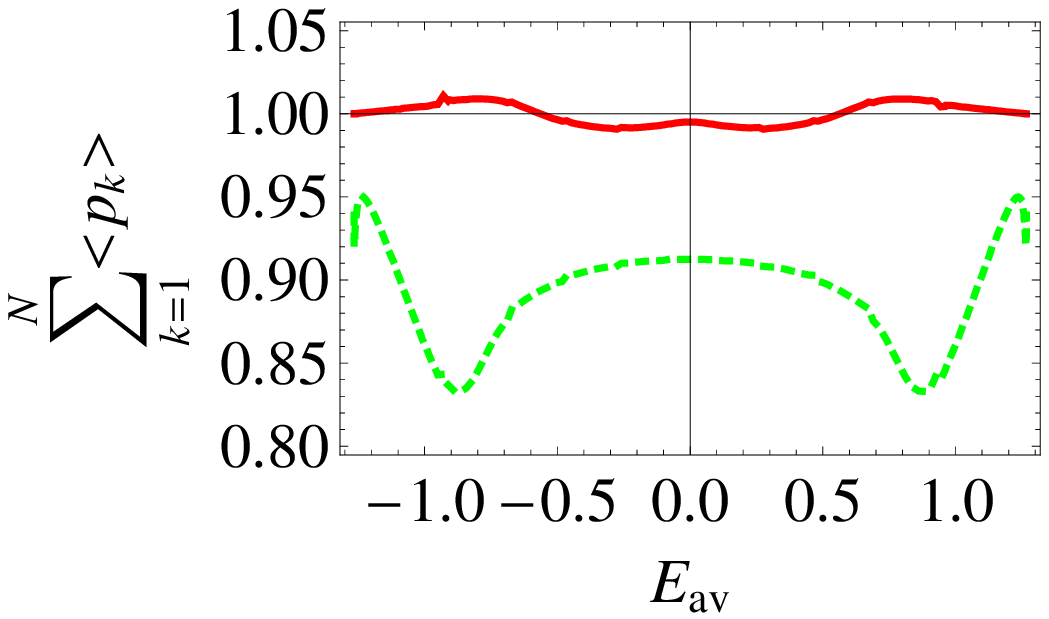} }
}
\end{pspicture}
\caption{Main figure: Function $\lambda[E_{\hbox{\scriptsize av}}]$ in various approximations for $N=12$ quantum spectrum represented by vertical lines. Long-dash magenta line --- small-$p_k$ approximation based on Eqs.(\ref{Vkpk1},\ref{pav2}); short-dash green line --- mean-$\lambda$ approximation based on Eq.(\ref{Vkpk2}); thick solid red line --- finite-$N$ corrected calculation based on Eq.(\ref{Vkpk2A}) and the ansatz of Section~\ref{ansatz}; thin solid black line --- asymptotic behavior in the large-$N$ limit given by Eqs.(\ref{lambdaA},\ref{lambdaB}).  Inset: Normalization test of the mean-$\lambda$ approximation (dashed green line) and finite-$N$ corrected calculation (solid red line).}
\end{figure}

Figure~1 presents curves for $\lambda[E_{\hbox{\scriptsize av}}]$  obtained in various theoretical approximations: namely, small-$p_k$ approximation based on Eqs.(\ref{Vkpk1},\ref{pav2}); mean-$\lambda$ approximation based on Eq.(\ref{Vkpk2}); and the calculation with finite-$N$ corrections based on Eq.(\ref{Vkpk2A}) and the ansatz of Section~\ref{ansatz}. The numerical algorithms for computing the latter two curves are described in \ref{algorithmsL}. Figure~1 also includes curves representing the large-$N$ asymptotic behavior (\ref{lambdaA},\ref{lambdaB}). 

The calculations of all three curves are based on satisfying energy constraint (\ref{epsavav}). The small-$p_k$ approximation, guarantees that the normalization constraint (\ref{normav}) is satisfied at the same time. However, the other two approximations do not guarantee it {\it a priori}. Therefore, the check of how the normalization constraint is satisfied constitutes an indication of the accuracy of the approximations. This check is presented in the inset of Fig.~1 as a function of $E_{\hbox{\scriptsize av}}$. It indicates that the normalization for the mean-$\lambda$ approximation falls short of 1 by about 10 per cent (roughly $1/N$). The normalization for the finite-$N$ corrected calculation deviates from 1 within $\pm 1$ per cent. 

\begin{figure}[p] \setlength{\unitlength}{1.0mm}

\psset{xunit=0.1cm,yunit=0.1cm}
\begin{pspicture}(0,0)(157,190)
{ 

\put(0.7,17.5){ \textsf{\Large (a)} }
\put(5.9,17.5){ \textsf{\Large (e)} }
\put(11.1,17.5){ \textsf{\Large (i)} }

\put(0,13.5){ \epsfxsize= 1.9in \epsfbox{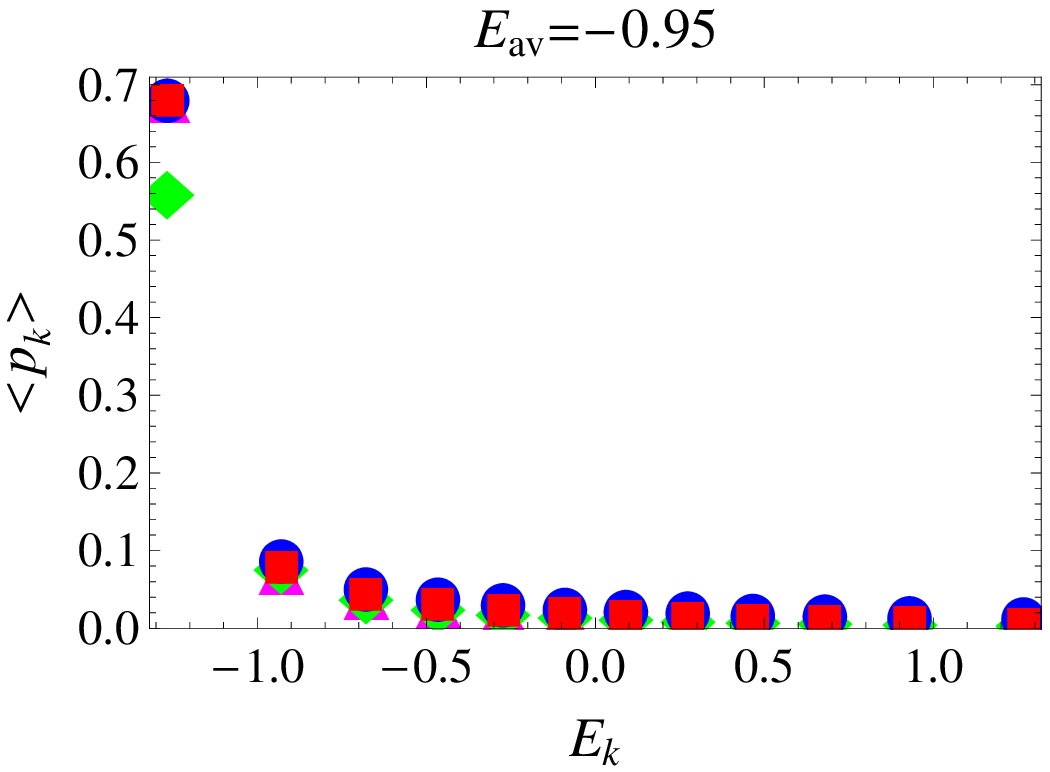} }
\put(5.2, 13.5){ \epsfxsize= 1.94in \epsfbox{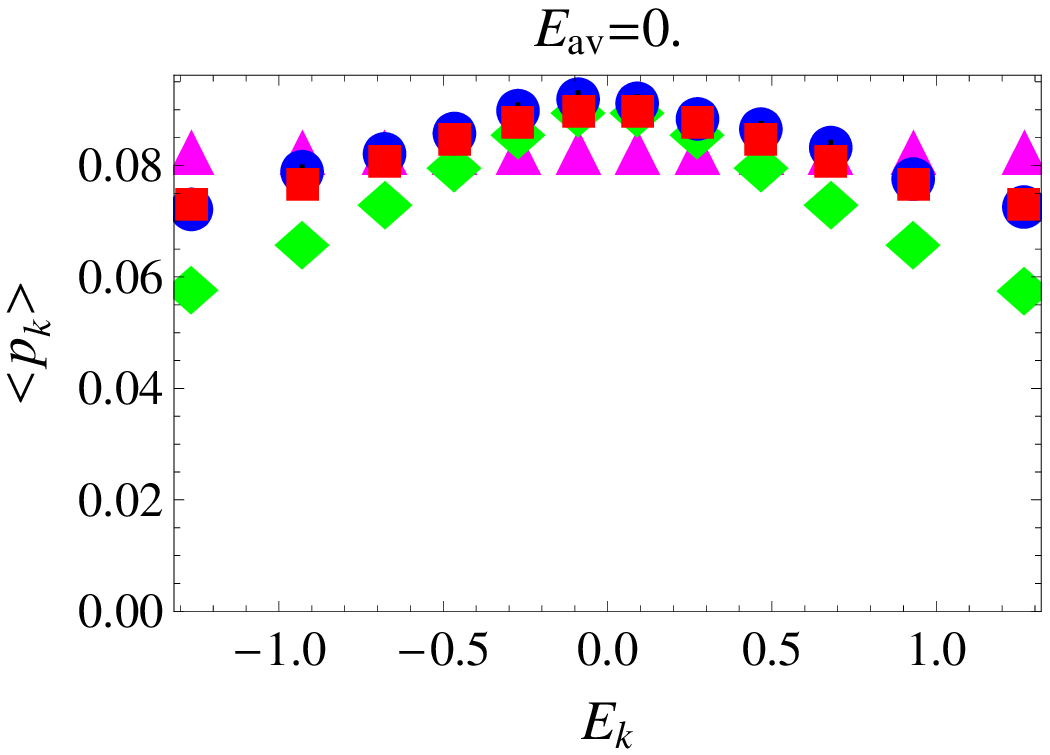} }
\put(10.4, 13.5){ \epsfxsize= 1.94in \epsfbox{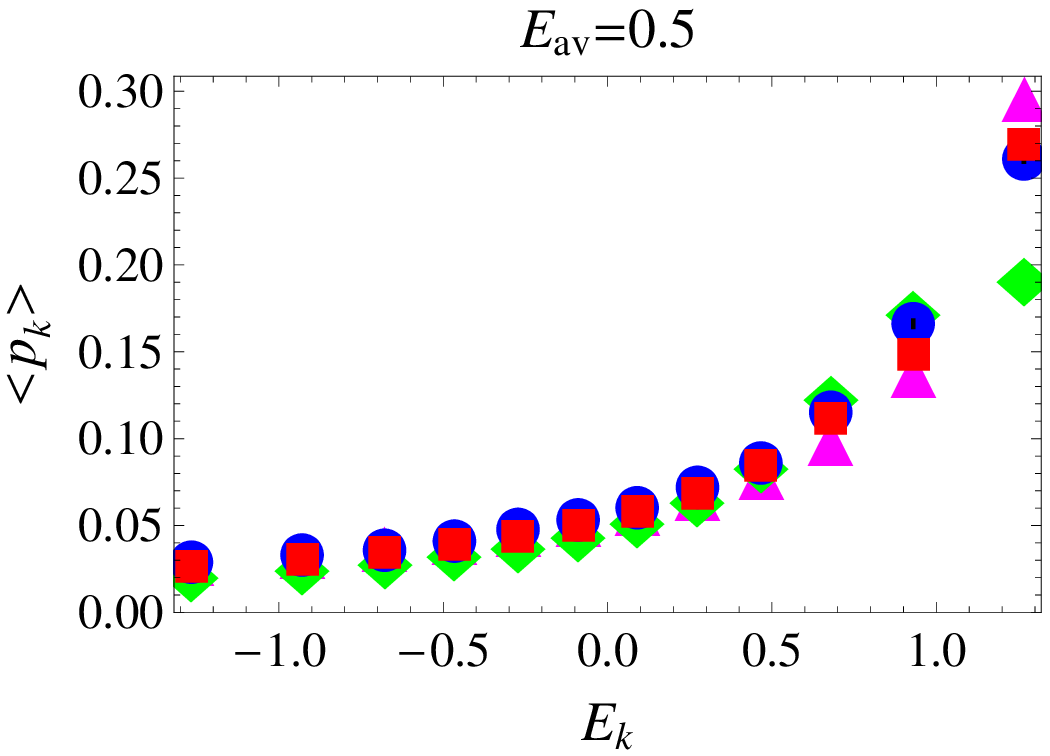} }

\put(0.7,13){ \textsf{\Large (b)} }
\put(5.9,13){ \textsf{\Large (f)} }
\put(11.4,13){ \textsf{\Large (j)} }

\put(0,9){ \epsfxsize= 1.92in \epsfbox{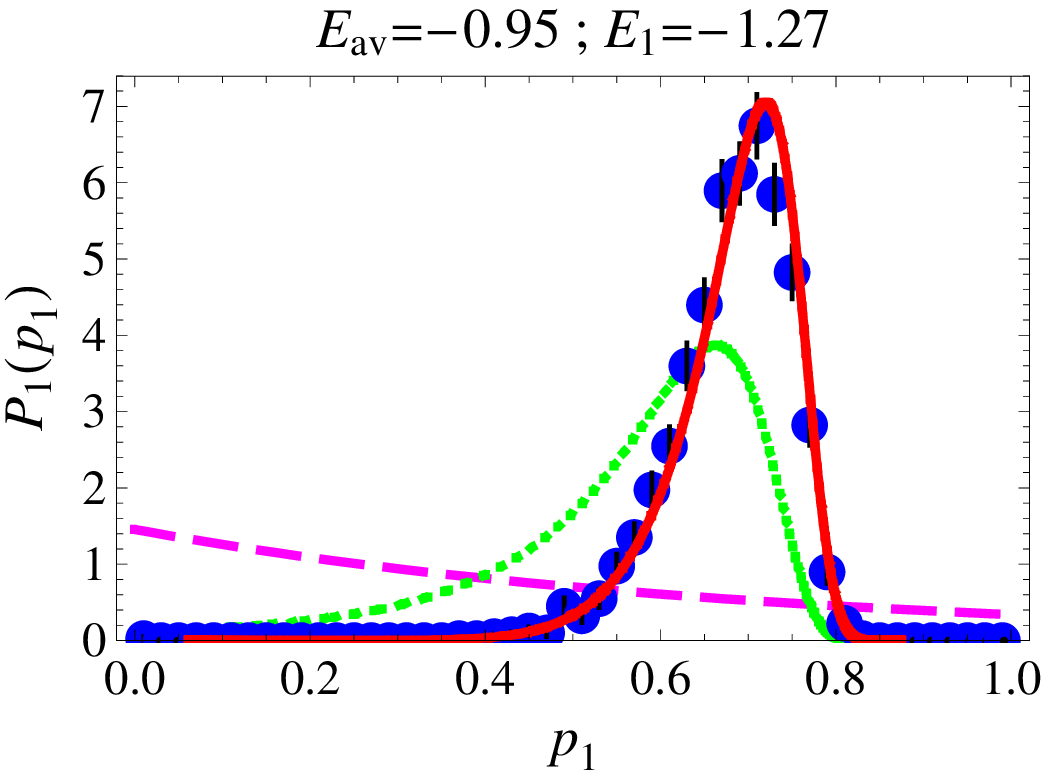} }
\put(5.2, 9){ \epsfxsize= 2in \epsfbox{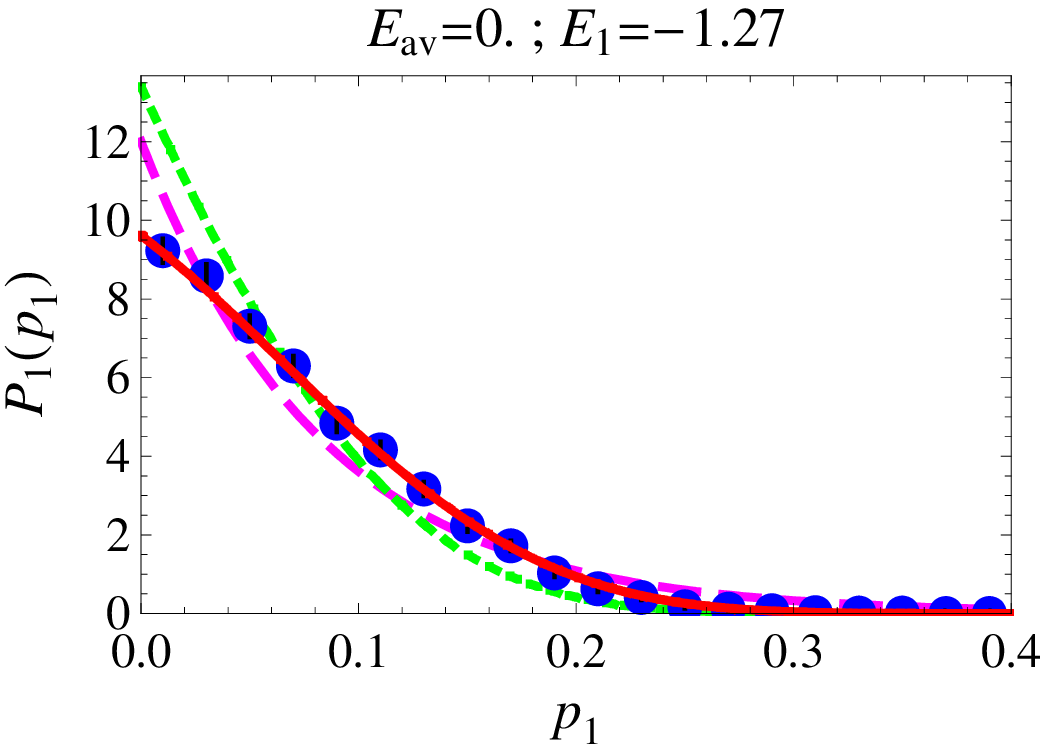} }
\put(10.4, 9){ \epsfxsize= 2in \epsfbox{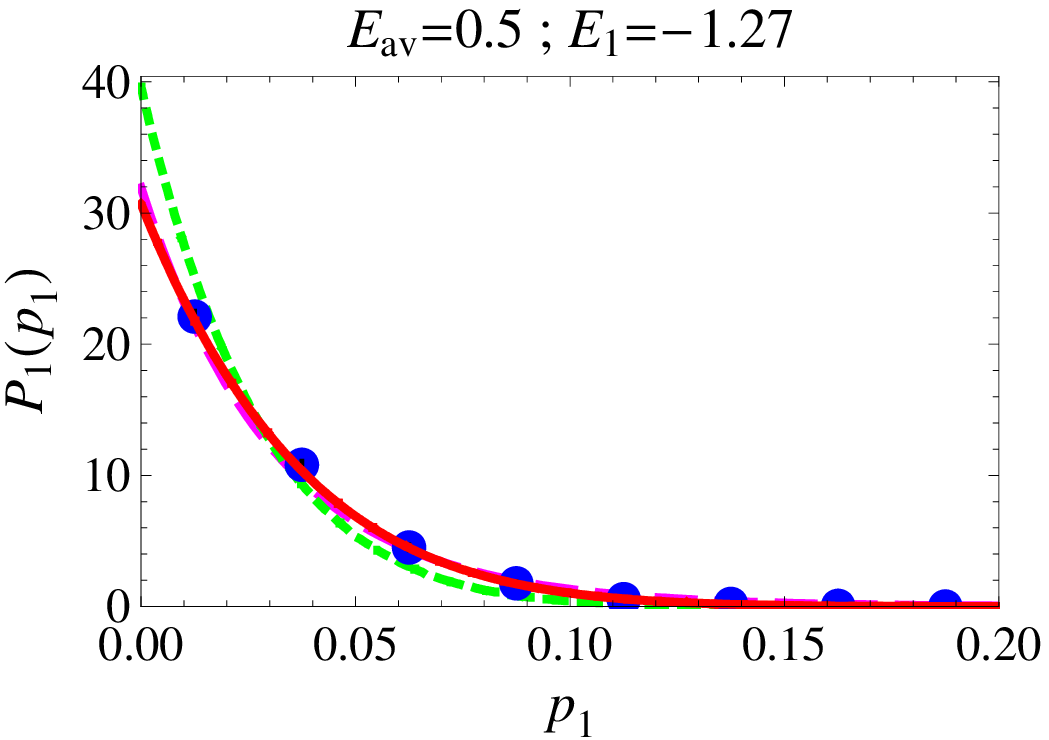} }

\put(0.7,8.5){ \textsf{\Large (c)} }
\put(5.9,8.5){ \textsf{\Large (g)} }
\put(11.4,8.5){ \textsf{\Large (k)} }

\put(0,4.5){ \epsfxsize= 2in \epsfbox{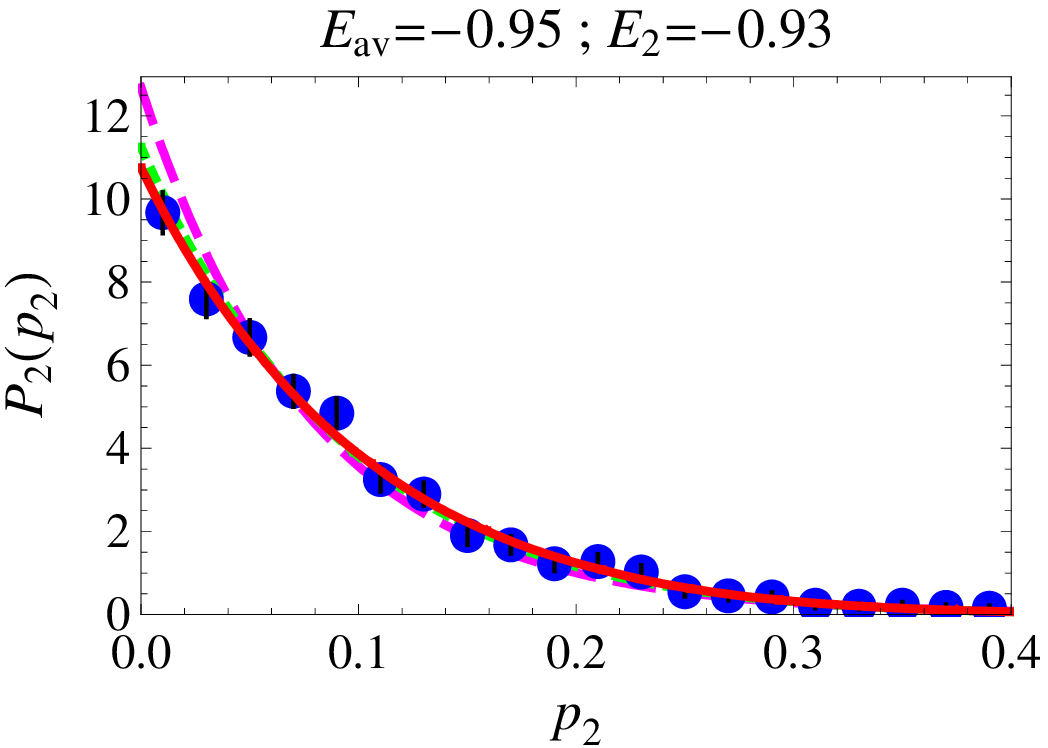} }
\put(5.2,4.5){ \epsfxsize= 2in \epsfbox{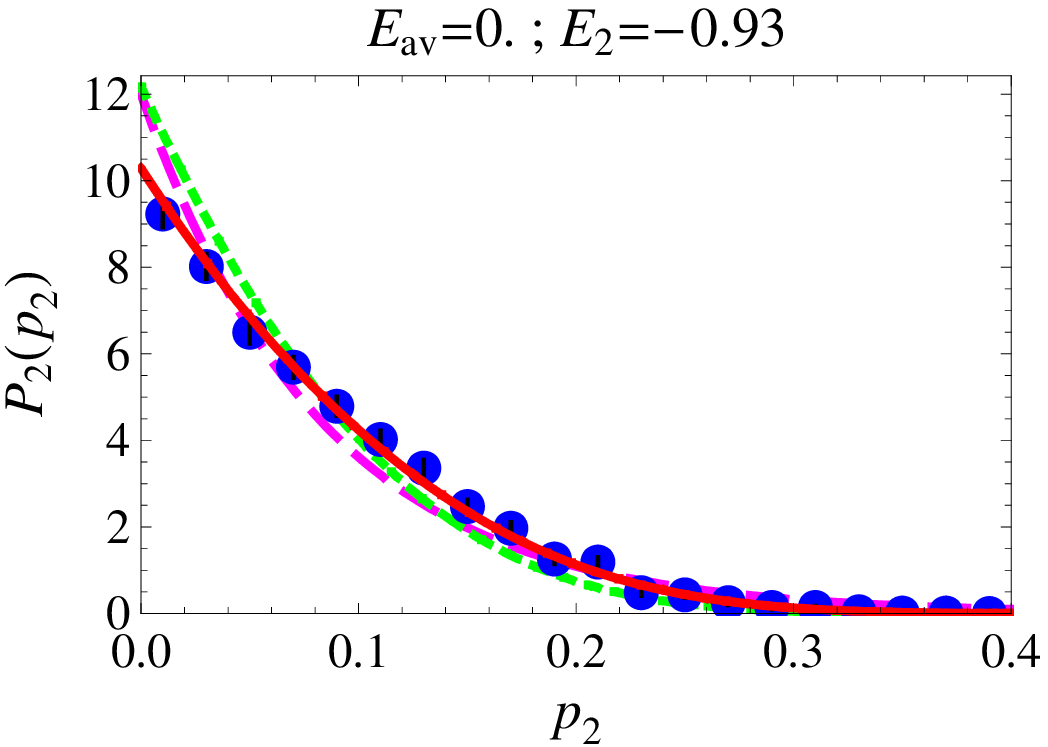} }
\put(10.4,4.5){ \epsfxsize= 2in \epsfbox{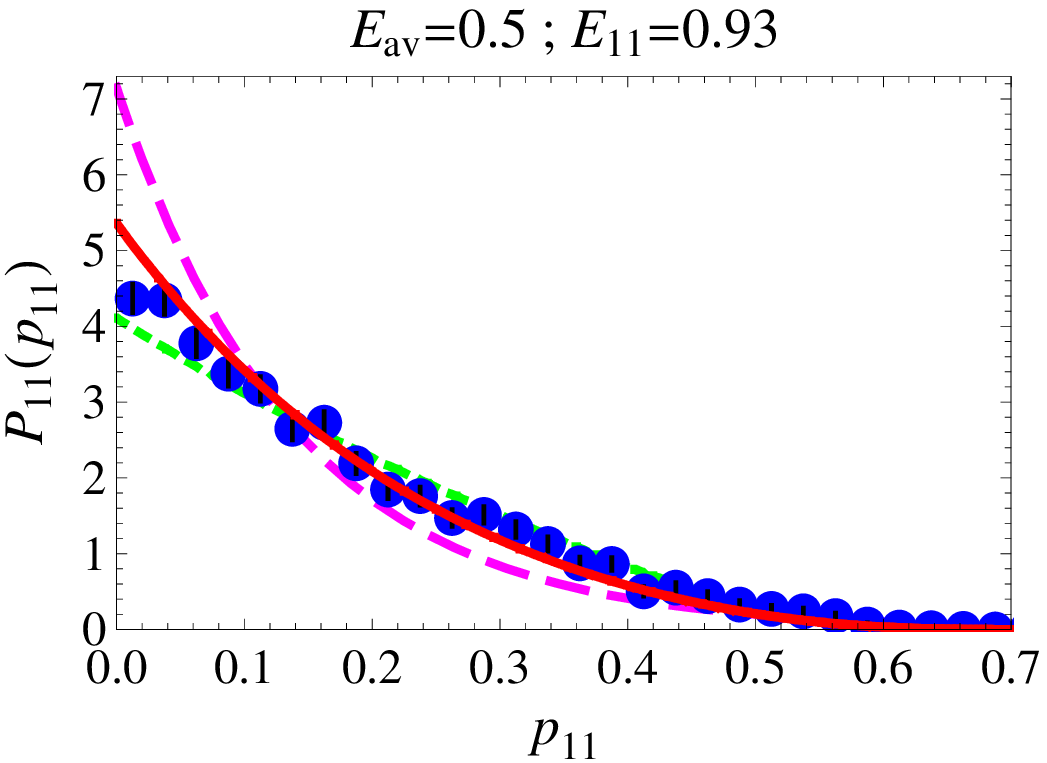} }

\put(0.7,4){ \textsf{\Large (d)} }
\put(5.9,4){ \textsf{\Large (h)} }
\put(11.4,4){ \textsf{\Large (l)} }

\put(0,0){ \epsfxsize= 2in \epsfbox{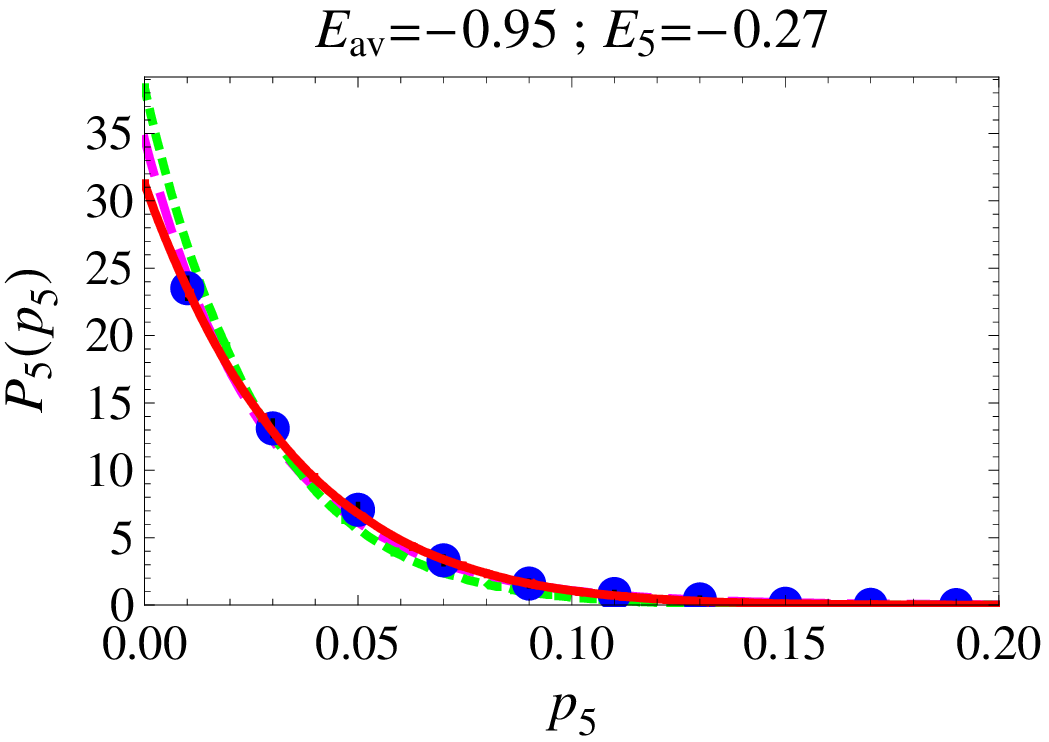} }
\put(5.2,0){ \epsfxsize= 2in \epsfbox{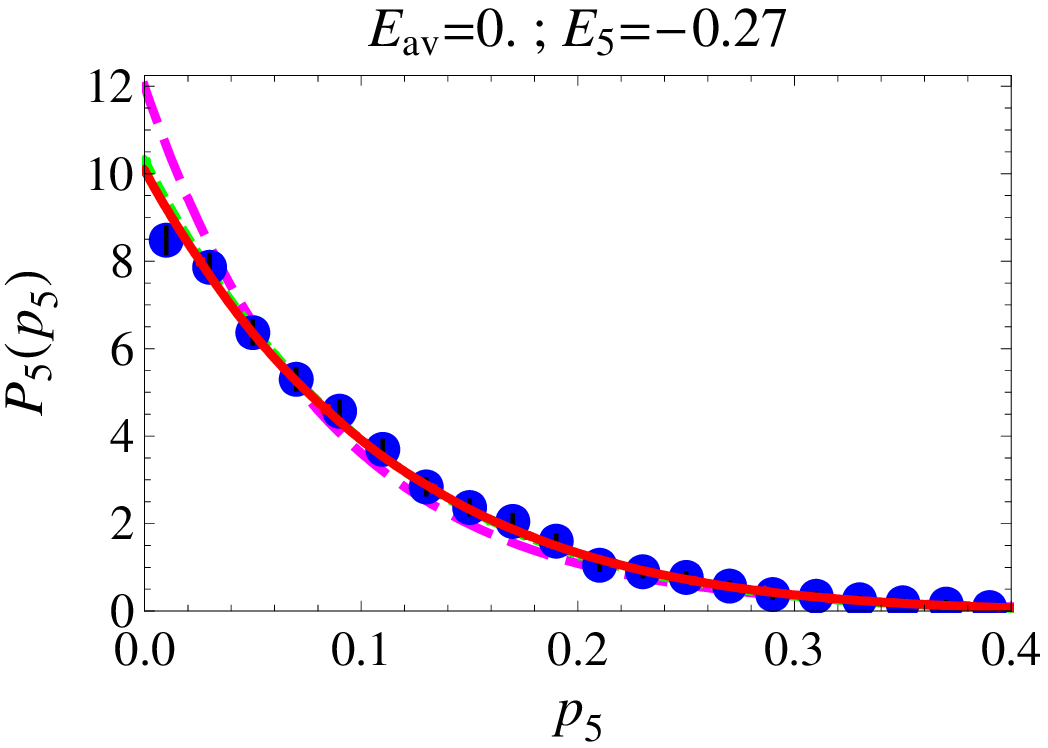} }
\put(10.4,0){ \epsfxsize= 2in \epsfbox{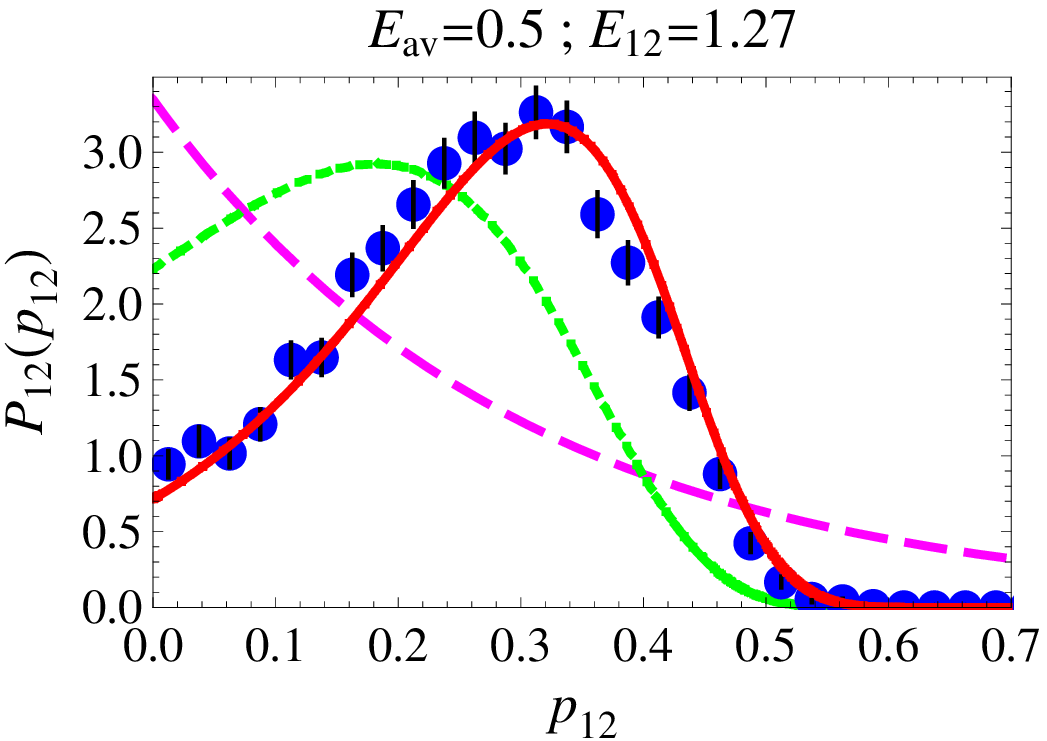} }

}
\end{pspicture} 
\caption{Comparison between the results of direct Monte-Carlo sampling of the QMC ensemble for $N=12$ quantum spectrum given in the text and in Fig.~1 (blue circles with statistical error bars, when visible) and three analytical approximations based on functions $\lambda[E_{\hbox{\scriptsize av}}]$ presented in Fig.~1: magenta triangles and magenta long-dash lines --- small-$p_k$ approximation based on Eqs.(\ref{Vkpk1},\ref{pav2}); green diamonds and green small-dash line --- mean-$\lambda$ approximation based on Eq.(\ref{Vkpk2});
red squares and red solid lines --- finite-$N$ corrected calculation based on Eq.(\ref{Vkpk2A}) and the ansatz of Section~\ref{ansatz}.}
\label{fig-direct} 
\end{figure}

Figure~2 presents detailed comparison between Monte-Carlo results and the above analytical approximations for three different values of $E_{\hbox{\scriptsize av}}$, $\{ -0.95, 0, 0.5 \}$.The numbers of Monte Carlo points accepted by the algorithm from \ref{algorithmMC} for the three values of $E_{\hbox{\scriptsize av}}$ were, respectively,  2000, 5000 and 5000.  The first row of frames in Fig.~2 represents the comparison for the values of the average occupation numbers $\langle p_k \rangle$  corresponding to twelve energies $E_k$ in the spectrum. The remaining three rows represent the comparison for the three (out of twelve) probability distributions $P_k(p_k)$ for each of the above three average energies.  

In the case of $E_{\hbox{\scriptsize av}} = -0.95$, the Monte-Carlo result qualitatively corresponds to the prediction of Ref.\cite{Fine-09-statistics} that, for macroscopic systems (meaning Gaussian density of states in the large-$N$ limit) at realistic positive Hilbert space temperatures $1/(N\lambda)$ (i.e. with $E_{\hbox{\scriptsize av}} < E_{\hbox{\scriptsize av}0}$), the QMC ensemble leads to condensation into the lowest energy state with $P_1(p_1)$ having the character of $\delta$-function peaked at $\langle p_1 \rangle$, while all other states have small average occupations $\langle p_k \rangle$ and exhibit exponential shapes of $P_k(p_k)$. What one sees in Figs.~2(a,b,c,d) is a finite-$N$ realization of the above condensation. The Monte-Carlo results indicate  that all $P_k(p_k)$ except for $P_1(p_1)$ corresponding to the lowest spectral level 
$E_1 = -1.27$ have nearly exponential shape well describable by all three of the approximations considered [Fig.~2(c,d)]. The differences, however, become pronounced once $P_1(p_1)$  is considered. Figure~2(b) indicates, that the small-$p_k$ approximation fails completely to describe $P_1(p_1)$, the mean $\lambda$ approximation reproduces the Monte-Carlo results qualitatively, but with large quantitative discrepancy, while the finite-$N$ corrected calculation is in a good quantitative agreement with the Monte-Carlo results.

Despite failing to predict $P_1(p_1)$, the small-$p_k$ approximation predicts accurately the value of $\langle p_1 \rangle$ in Fig~2(a). As discussed in Ref.\cite{Fine-09-statistics}, it is the consequence of the fact that this approximation accurately describes all $P_k(p_k)$ and $\langle p_k \rangle$ for $k > 1$, but then, since it automatically satisfies the normalization condition (\ref{normav}), it guarantees that the value of $\langle p_1 \rangle$ is also accurate.

The example of $E_{\hbox{\scriptsize av}} = 0.5$ corresponds to the negative Hilbert space temperature $1/(N\lambda)$. In this case, the QMC ensemble tends to condense to the highest energy state. The facts, that, in this case, the distribution of $P_{12}(p_{12})$ [Fig.~2(l)] is rather broad, and that the form of $P_{11}(p_{11})$ [Fig.~2(k)] for the second highest level also noticeably deviates from exponential, are the finite-$N$ effects.
Here again, the small-$p_k$ approximation fails to describe $P_{11}(p_{11})$ and $P_{12}(p_{12})$ , the mean-$\lambda$ calculation gives correct qualitative trends with quantitative errors, and the finite-$N$-corrected calculation exhibits a good quantitative agreement with the Monte-Carlo results. For $k<10$ all three approximations give nearly exponential decays for  $P_k(p_k)$, in good agreement with the  Monte-Carlo results [see, e.g., Fig.~2(j)]. Since the small-$p_k$ approximation fails for three highest levels, it is no longer guaranteed to produce the accurate values of $\langle p_{10} \rangle$, $\langle p_{11} \rangle$ and $\langle p_{12} \rangle$ and, in fact, the noticeable discrepancy can be seen in Fig.~2(i).

The example of $E_{\hbox{\scriptsize av}} = 0$ corresponds to the infinite Hilbert space temperature. In this case, the small-$p_k$ approximation predicts that all $P_k(p_k)$ have the same exponential shape, and that, independently of $E_k$, $\langle p_k \rangle = 1/N$. This would indeed be the case in the limit $N \rightarrow \infty$. However, here the symmetric position of $E_{\hbox{\scriptsize av}} = 0$ with respect to the quantum spectrum leads to a more systematic character of finite-$N$ corrections, which, in turn, leads to a larger discrepancy with the Monte-Carlo results for all $k$. The origin of this systematic discrepancy is in the corrections to the linearized version of integral in Eq.(\ref{Vkpk2}). The linearization of that integral with respect of $p_k$ leads to the small-$p_k$ approximation. The higher order corrections are, in general, partly canceled, because $\lambda[E]$ in the integral changes sign, when the integration range includes $E=0$ somewhere in the middle. However, when $E_{\hbox{\scriptsize av}} = 0$, this sign-change point coincides with the lower integration limit. As a result, $\lambda[E]$ does not change sign within the integration range for any $k$, which leads to the above larger discrepancy. Two other approximations capture the above systematic trend, but here again, the finite-$N$ corrected calculation exhibits a better quantitative agreement with the Monte-Carlo results.  

It should be emphasized that the finite-$N$ corrected calculation is based on an ansatz of Section~\ref{ansatz}, which, although meaningful and apparently very effective, does not amount to an exact calculation.  Therefore, it is expected to show small deviations from the Monte-Carlo results, which can, indeed, be seen in Figs.2(k,l).

The statistics of eigenstate participation for the entire isolated system reported in this section is not a direct test of the presence or the absence of BG equilibrium for a small subsystem. However, in this respect, the results of significance are those shown in Figs.~2(a,e,i). They represent, in a sense, the participation functions for conventional micro-canonical ensembles with different positions of micro-canonical energy windows. These participation functions have  algebraic dependence on energy mostly captured by the small-$p_k$ result (\ref{pav2}). This slow non-exponential dependence precludes one from treating the participation of eigenstates as a conventional canonical ensemble. As a result,  the broad participation of eigenstates in the QMC ensemble implies a broad mixture of conventional thermal states, which, for a small subsystem, does not result in a single thermal BG distribution --- the conclusion confirmed by direct calculation in Ref.\cite{Fine-09-statistics}. 
In the next section, we present numerical evidence corroborating this conclusion.

\section{Implications of the QMC ensemble for a subsystem within a small spin system}
\label{spins}

Reference \cite{Fine-09-statistics} derived analytical results for the density matrix of a small subsystem within a larger isolated quantum system described by the QMC ensemble. When the whole system is macroscopic (i.e. the number of levels $N$ is exponentially large), the QMC-based density matrix of the small subsystem is a weighed sum of two terms:  the zero-temperature density matrix and the infinite temperature density matrix.  In the finite-$N$ case, one has to solve a system of equations obtained in \cite{Fine-09-statistics}  --- not to be attempted in the present work. Instead, in this section, we demonstrate numerically that the following qualitative property of the macroscopic limit also appears in the finite-$N$ case. Namely, for a subsystem within an isolated quantum system, the lowest energy state and the states on the high-energy end of the spectrum have larger occupations when the isolated system is sampled according to the QMC ensemble than when it is sampled according to the conventional canonical ensemble. Correspondingly, the intermediate energy states have smaller occupations for the QMC ensemble. 

Below we consider a 12-level system consisting of two spins 1/2 and spin 1, and examine numerically the implications of the QMC ensemble for the subsystem of two spins 1/2. 

As discussed in the previous section, the 12-level limitation is due to the computational difficulty of sampling high-dimensional Hilbert spaces. In principle, in order to exhibit deviations from the BG statistics, the subsystem should have at least three levels, and therefore one could think that it is more natural to consider the spin 1 as a subsystem.  However, with only 3 levels in the subsystem the differences between the QMC-based results and the BG statistics would be less instructive.

\begin{figure}[t] \setlength{\unitlength}{1.0mm}

\psset{xunit=0.1cm,yunit=0.1cm}
\begin{pspicture}(0,0)(157,25)
{ 
\put(3,0){ \epsfxsize= 4in \epsfbox{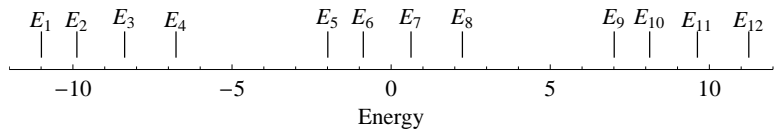} }
}
\end{pspicture} 
\caption{Energy spectrum of the 12 level system (two spins 1/2 plus spin 1) considered in Section~\ref{spins}}
\label{fig-spectrum} 
\end{figure}

We assume the the environment represented by spin 1 does not interact with the subsystem of two coupled spins 1/2. Hence the Hamiltonian of the entire system is:
\begin{equation}
{\cal H} = {\cal H}_I + {\cal H}_S,
\label{H}
\end{equation} 
where 
\begin{equation}
{\cal H}_I = \gamma_I H (I_{1z} + I_{2z}) + J_x I_{1x}  I_{2x} + J_y I_{1y}  I_{2y} + J_z I_{1z}  I_{2z}
\label{HI}
\end{equation}
and 
\begin{equation}
{\cal H}_S = \gamma_S H S_z 
\label{HS}
\end{equation}
are, respectively, the Hamiltonians for the two spins 1/2 and for spin 1. Here $I_{m \alpha}$ are the operators of for the $\alpha$-components ($\alpha = x,y,z$) of the two spins 1/2 ($m=1,2$),  $S_z$ is the operator of the $z$-component of spin 1, $H=3$ is the external magnetic field, $\gamma_I = 0.7$ and $\gamma_S = 3$ are the gyromagnetic ratios for spins 1/2 and for spin 1, respectively; $J_x = -2$, $J_y = -1$, $J_z = 0.5$ are anisotropic coupling constants. Since spin 1 plays the role of the environment, the parameters of the Hamiltonian were chosen such that the spread of the spin-1 levels is significantly broader than that of the two-spin-1/2 subsystem.  The energy spectrum of the Hamiltonian (\ref{H}) with the above choice of parameters is shown in Fig.~\ref{fig-spectrum}. It has three clearly identifiable 4-level groups.

\begin{figure}[t] \setlength{\unitlength}{1.0mm}

\psset{xunit=0.1cm,yunit=0.1cm}
\begin{pspicture}(0,0)(157,80)
{ 
\put(3,0){ \epsfxsize= 4.5in \epsfbox{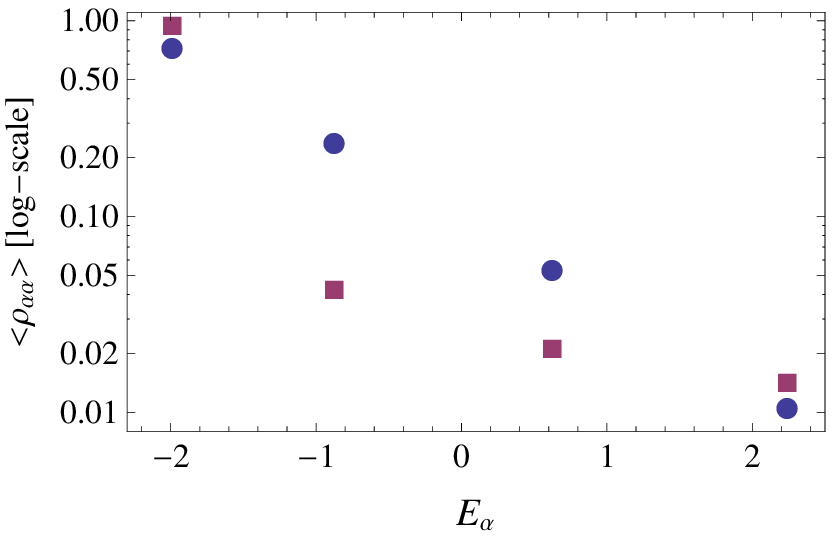} }
\put(9.3,3.8){ \epsfxsize= 1.9in \epsfbox{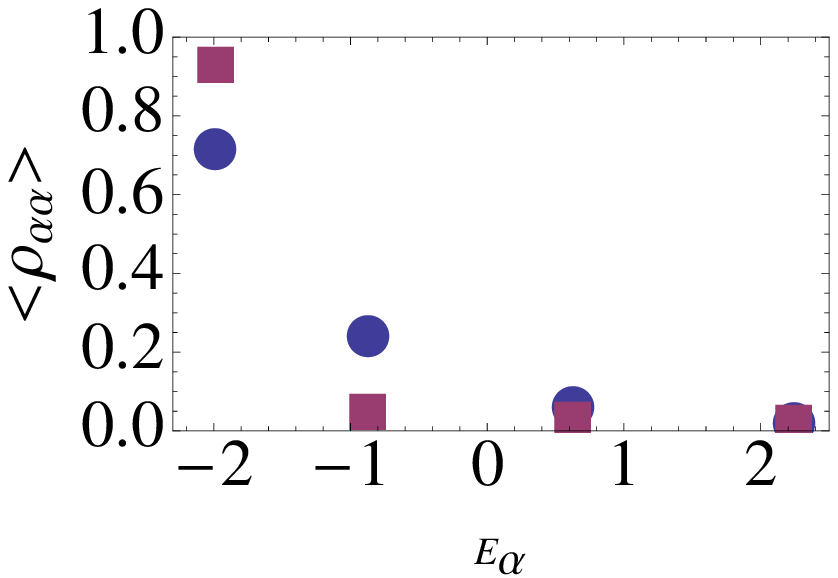} }
}
\end{pspicture} 
\caption{Diagonal elements of ensemble-averaged density matrix $<\rho_{\alpha \alpha}>$  for the subsystem of two spins 1/2 considered in Section~\ref{spins} in the basis of eigenstates of the Hamiltonian ${\cal H}_I$. The subsystem eigenstates are labeled by index $\alpha$ with corresponding energies denoted as $E_{\alpha}$. Magenta squares represent the QMC ensemble, blue circles represent the canonical ensemble. Inset: same results on the linear scale.} 
\label{fig-rho} 
\end{figure}

We compare the subsystem density matrices for the QMC ensemble and for the canonical ensemble at the same value of the average energy of the entire isolated system. In order to do this, we first fixed the temperature of the canonical ensemble T=1, then calculated the corresponding average energy $E_{\hbox{av}} = -10.55$, and then generated 2000 sample wave functions from the QMC ensemble for the above value of $E_{\hbox{av}}$. Afterwards, we calculated the average subsystem density matrices corresponding to the both ensembles. The results for the diagonal elements of the density matrix $\rho_{\alpha \alpha}$ in the eigenbasis of the Hamiltonian ${\cal H}_I$ are presented in Fig.~\ref{fig-rho}.
As expected from the large-$N$ limit\cite{Fine-09-statistics}, the values of $\rho_{\alpha \alpha}$ for the QMC ensemble are larger than those for the canonical ensemble for the lowest and the highest subsystem energy levels, and correspondingly smaller for the middle two levels.

\begin{figure}[t] \setlength{\unitlength}{1.0mm}

\psset{xunit=0.1cm,yunit=0.1cm}
\begin{pspicture}(0,0)(157,80)
{ 
\put(3,0){ \epsfxsize= 4.5in \epsfbox{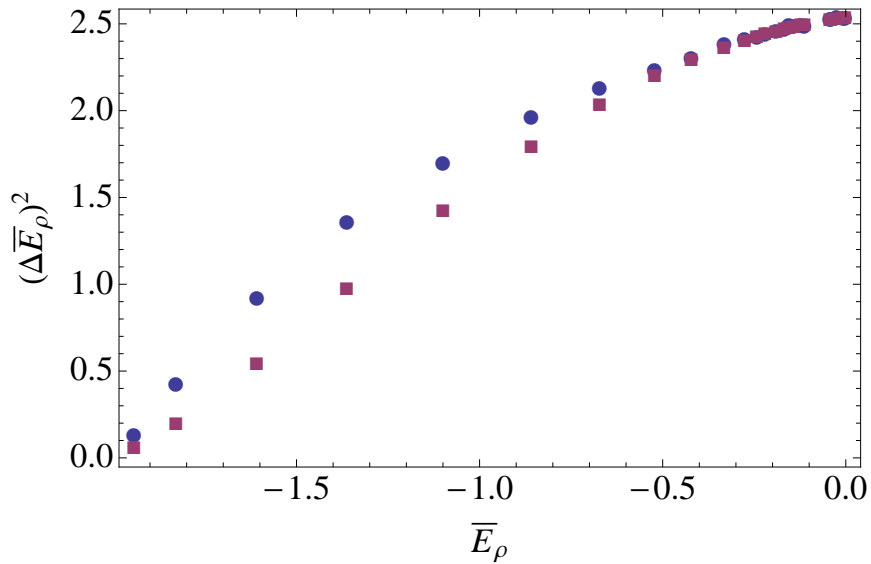} }
}
\end{pspicture} 
\caption{Energy variances  for the subsystem of two spins 1/2 given by Eq.(\ref{DE}) as a function of the average subsystem's energy.  Magenta squares represent the QMC ensemble, blue circles represent the canonical ensemble.} 
\label{fig-DE} 
\end{figure}

If one can have only limited experimental access to the properties of the subsystem, the QMC ensemble can be identified by looking at the variance of energies for the subsystem states:
\begin{equation}
(\Delta E_{\rho})^2 = \sum_{\alpha} \rho_{\alpha \alpha} (E_{\alpha} - \bar{E}_{\rho})^2,
\label{DE}
\end{equation}
where index $\alpha$ labels subsystem's eigenstates, $E_{\alpha}$ denotes the corresponding subsystem eigenenergies, and $\bar{E}_{\rho} = \sum_{\alpha} \rho_{\alpha \alpha} E_{\alpha}$ is the average energy of the subsystem.
In Fig.~\ref{fig-DE}, we plot subsystem energy variances for the QMC and the canonical ensembles as a function of $\bar{E}_{\rho}$.  For $E_{\hbox{av}} = E_1$ (see Fig.~\ref{fig-spectrum}) and $E_{\hbox{av}} = 0$, the predictions of the two ensembles are identical, and hence the variances are the same. (In the first case, only the ground state is occupied and $\bar{E}_{\rho} = \hbox{min}[E_{\alpha}]$, while in the second case all subsystem's states are equally occupied and  $\bar{E}_{\rho} = 0$.) In-between the energy variance for the QMC ensemble is always larger. This is true not only for the subsystem, but also for the entire system.

\begin{figure}[t] \setlength{\unitlength}{1.0mm}

\psset{xunit=0.1cm,yunit=0.1cm}
\begin{pspicture}(0,0)(157,90)
{ 
\put(3,0){ \epsfxsize= 4.5in \epsfbox{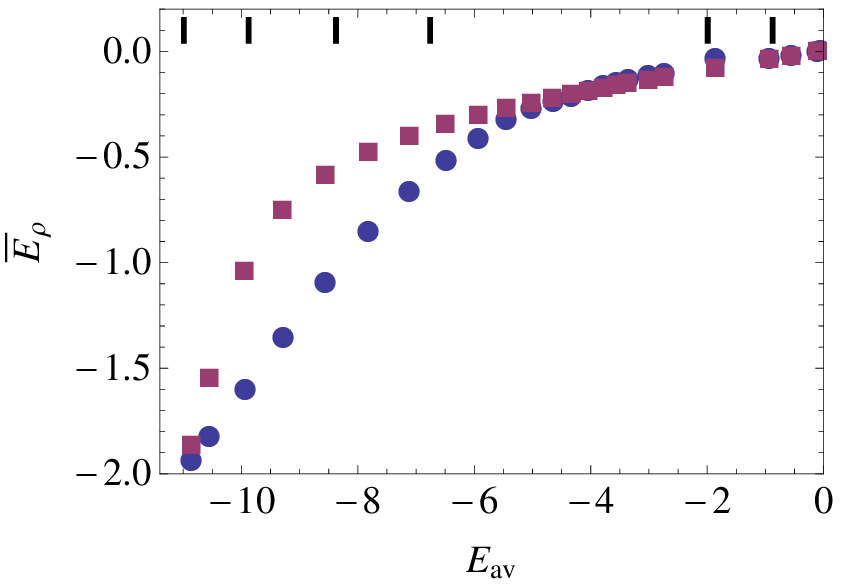} }
\put(8.8,1.8){ \epsfxsize= 1.9in \epsfbox{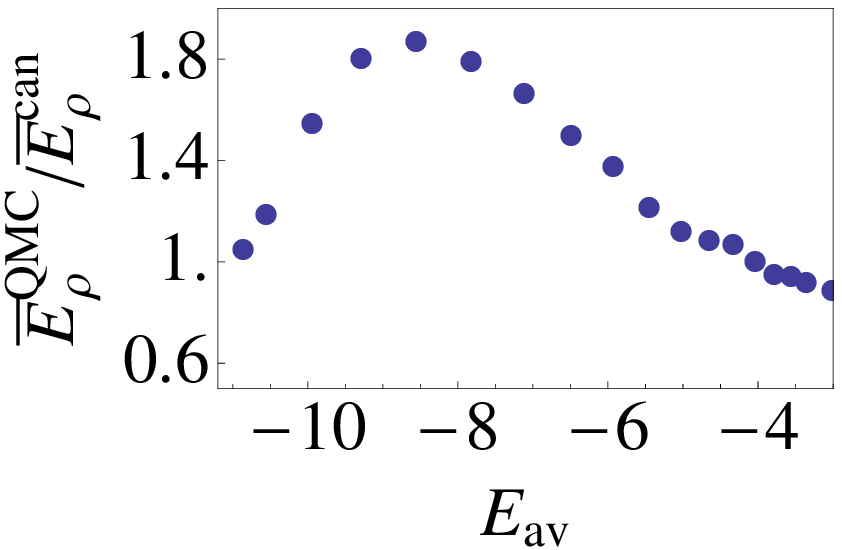} }
}
\end{pspicture} 
\caption{Average energy  for the subsystem of two spins 1/2 as a function of the average energy for the whole system.  Magenta squares represent the QMC ensemble, blue circles represent the canonical ensemble. Inset: Ratio of the subsystem energy for the QMC over the subsystem energy for the canonical ensemble.} 
\label{fig-Ebar} 
\end{figure}

An interesting fact is that for the parameters considered, the subsystem average energies for the QMC and the canonical ensembles are different even though the energy of the entire system is the same for the both ensembles. The subsystem energies for the two ensembles are, respectively, $\bar{E}^{\hbox{\scriptsize QMC}}_{\rho} = -1.83$ and 
$\bar{E}^{\hbox{\scriptsize can}}_{\rho} = -1.55$, i.e., in the QMC case,  the subsystem is in a sense ``colder".  We have investigated this issue further, and in Fig.~\ref{fig-Ebar} have plotted the result for $\bar{E}_{\rho}$ for the both ensembles as a function of $E_{\hbox{\scriptsize av}}$ of the total system. Our findings indicate that there exists a critical value $E^{\hbox{\scriptsize cr}}_{\hbox{\scriptsize av}} \approx -4$ such that 
$\bar{E}^{\hbox{\scriptsize QMC}}_{\rho} < \bar{E}^{\hbox{\scriptsize can}}_{\rho} $ for $E_{\hbox{\scriptsize av}} < E^{\hbox{\scriptsize cr}}_{\hbox{\scriptsize av}} $, and $\bar{E}^{\hbox{\scriptsize QMC}}_{\rho} > \bar{E}^{\hbox{\scriptsize can}}_{\rho} $ for
$E^{\hbox{\scriptsize cr}}_{\hbox{\scriptsize av}}  < E_{\hbox{\scriptsize av}} <0$. However for $E^{\hbox{\scriptsize cr}}_{\hbox{\scriptsize av}}  < E_{\hbox{\scriptsize av}} <0$, the results for the two ensembles are very close (the high-temperature limit) and hence the difference between $\bar{E}^{\hbox{\scriptsize QMC}}_{\rho}$ and $\bar{E}^{\hbox{\scriptsize can}}_{\rho}$ is insignificant and difficult to resolve statistically.

The difference between $\bar{E}^{\hbox{\scriptsize QMC}}_{\rho}$ and $\bar{E}^{\hbox{\scriptsize can}}_{\rho}$ may, perhaps, be used as another experimentally identifiable signature of the QMC ensemble. It should be noted, however, that the above difference can only appear when the subsystem particles are different from the environment particles. The subsystem energies for the two ensembles are guaranteed to be the same, when, for example, the whole system is describable by a translationally invariant Hamiltonian of spins 1/2, and the interaction between the subsystem of a few spins 1/2 and the environment is negligible.

\section{Prospects}
\label{prospects}

Here we review relevant open questions.

\subsection{Macroscopic systems}

It appears now very likely that the micro-canonical narrow energy window condition is crucial for the justification of BG equilibrium in macroscopic systems. What, however, can justify this condition itself? The only argument known to us is associated with quantum collapse --- also referred to as quantum measurement or quantum projection\cite{Fine-09-statistics,Deutsch-09}. Eigenstates with significantly different energy values should have macroscopically distinguishable   characteristics. Hence the mixture of macroscopically distinguishable states, coherent or not, would not be tolerated in nature, and instead the mixture would collapse to one macroscopic state with usual probabilities predicted by quantum mechanics.

The disturbing part of the above argument is that the quantum collapse is an essentially non-linear process, and hence cannot appear in the quantum-mechanical theory in the course of linear evolution of a wave function or a density matrix. Therefore, it implies that BG equilibrium cannot be justified in the framework of linear quantum mechanics --- it requires a non-linear quantum collapse into the narrow micro-canonical energy window.

One can still ask a different question: If the quantum collapse occurred only once in the evolution of a macroscopic system, is it realistic to expect that external perturbations on otherwise isolated macroscopic quantum system would force the spread of participating eigenstates over an energy window violating condition (\ref{T})? Is it further realistic to expect that the QMC ensemble can be realized in an isolated macroscopic quantum system that has once experienced quantum collapse? 
For a truly macroscopic system consisting of $10^{23}$ weakly interacting parts, the answer to both questions appears to be ``extremely unlikely''. The reasoning here is based on the central limit theorem: If such a system is subjected even to a drastic external perturbation, e.g. explosion, it is still likely that different parts of such a system, while being strongly perturbed, will continue to interact weakly, end hence the resulting total energy distribution will have Gaussian form, which would be sufficiently narrow to satisfy condition (\ref{T}).(See also Refs.\cite{Deutsch-91,Tasaki-98,Reimann-08}.) 

Yet a single perturbation makes the energy window of participating eigenstates somewhat broader. The next level of this argument would be to ask, how many sequential perturbations are necessary to force an isolated macroscopic system to violate condition (\ref{T}). This question requires further investigation. Realistic macroscopic systems are continuously subjected to the fluctuations of external potentials of electromagnetic or gravitational origin. Can these fluctuations drive an isolated system out of the micro-canonical energy window? If they can, but we do not observe any manifestation of it, does it mean that quantum collapse to the narrow energy window occurs continuously? If this happens, does it imply observable energy fluctuations not explainable by linear quantum mechanics?

\subsection{Small quantum systems with many levels}

The second line of investigation is related to dealing with quantum systems that do not yet have large number of degrees of freedom, but already have a very large number of quantum levels, e.g. $10\times 10 \times 10$ clusters of interacting spins 1/2 or a few cold atoms in an optical trap.

In this case, it appears realistic to isolate these systems against energy dissipation and, possibly, against decoherence. The central-limit-theorem-based considerations are not applicable to these systems. Therefore, it is clear that such systems can be driven out of condition described by conventional micro-canonical or canonical ensembles. Will the QMC ensemble generically emerge in this case, possibly, after several perturbations\cite{Ji-10}? If such an ensemble emerges, what are the observable signatures of it, e.g. in terms of measurable single particle properties? What would be the effects of decoherence and subsequent quantum collapse in this case? It is, in particular, possible that even if the broad QMC energy window collapses towards the canonical or micro-canonical shape, then the QMC-based calculations would predict the fluctuations of the resulting temperature. These questions require numerical investigations beyond the scope of this work. It is, in fact, likely that non-micro-canonical ensembles, and possibly QMC, routinely occur in the numerical studies of finite quantum systems.

\section{Conclusions}

In conclusion, we have demonstrated that the large-$N$ analytical description of the QMC ensemble developed in Ref.\cite{Fine-09-statistics} allows one to make good predictions for the Monte-Carlo sampling of the QMC ensemble for finite-$N$ systems. The description of Ref.\cite{Fine-09-statistics} is amenable to finite-$N$ corrections, and once these corrections are introduced, the theory produces accurate quantitative agreement with the Monte-Carlo results for systems with $N\sim 10$. 

We have also studied numerically the implications of the QMC ensemble for a subsystem of a small spin system. Our results indicate that, already in this rather artificial case, the behavior of subsystem's density matrix qualitatively follows the analytical result for the large-$N$ case. We have further suggested that the subsystem energy variance, as well as the average value of subsystem's energy itself,  can be used to discriminate experimentally between the QMC and  the BG statistics. 

In the context of foundations of quantum statistical physics, the apparent impossibility of obtaining the Boltzmann-Gibbs equilibrium from the QMC ensemble indicates that one should be looking closer at the basis and the applicability limit of the narrow energy window condition of the conventional micro-canonical ensemble. In this respect, quantum collapse may play a fundamental role.  

{\it Note added:} As we were finishing editing this manuscript for resubmission, we discovered a series of papers by Fresch and Moro\cite{Fresch-09,Fresch-10A,Fresch-10B} very closely related to the scope of the present work and Ref.\cite{Fine-09-statistics}.  In particular, in Ref.\cite{Fresch-09}, the above authors have pursued similar Monte-Carlo sampling of the QMC ensemble in finite dimensional Hilbert spaces. Their results appear to be consistent with and, in some aspects, complementary to ours. In particular, they have used Metropoilis-Hastings algorithm, as opposed to the direct sampling routine of our paper, and thereby were able to explore significantly larger Hilbert spaces. On the other hand, the present paper advances further the subject of finite-$N$ corrections to the large-$N$ QMC results. We have also learned from Ref.\cite{Fresch-09} that, already in 1990, Wootters have considered the QMC ensemble and arrived to what we call the small-$p_k$ approximation\cite{Wootters-90}.

\appendix

\section{Algorithms for finding $\lambda[E_{\hbox{\scriptsize av}}]$}
\label{algorithmsL}

Here we describe the algorithm for finding $\lambda[E_{\hbox{\scriptsize av}}]$ from the mean-$\lambda$ integral equation (\ref{Vkpk2}) --- ``Algorithm~A'', and the modification of Algorithm~A  including finite-$N$ corrections based on Eq.(\ref{Vkpk2A}) incorporating ansatz from Section~\ref{ansatz} --- ``Algorithm~B''.

Function $\lambda[E_{\hbox{\scriptsize av}}]$ is defined in the interval $[E_{\hbox{\scriptsize min}},E_{\hbox{\scriptsize max}} ]$. We use a non-uniform discretization of this interval with 228 grid points --- less dense in the middle and more dense near $E_{\hbox{\scriptsize min}}$ and $E_{\hbox{\scriptsize max}}$, were $\lambda[E_{\hbox{\scriptsize av}}]$ tends to diverge.
This near-divergence is, however, manageable, because, it is compensated by the overall exponential decrease of $V_k(p_k)$ and affects only the range of $p_k$, where $V_k(p_k)$ is exponentially small.

{\it Algorithm A: Outline} 

The algorithm first finds $\lambda[E_{\hbox{\scriptsize av}}]$ in the small-$p_k$ approximation on the basis of Eqs.(\ref{pav2}) and (\ref{epsavav}) and then iteratively improves it by (i) calculating $V_k(p_k)$ by substituting $\lambda[E_{\hbox{\scriptsize av}}]$ into Eq.(\ref{Vkpk2}), (ii) finding the corresponding $\langle p_k \rangle$, (iii) checking averaged energy condition (\ref{epsavav}); (iv) improving the value $\lambda[E_{\hbox{\scriptsize av}}]$ by matching condition (\ref{epsavav}) and then returning to step (i). As explained below, step (iv) is done ``locally'', i.e., for every $E_{\hbox{\scriptsize av}}$, on the grid it improves $\lambda[E_{\hbox{\scriptsize av}}]$ independently of the values of $\lambda[E_{\hbox{\scriptsize av}}]$ at other grid points, but then once all grid values of $\lambda[E_{\hbox{\scriptsize av}}]$ are improved, the algorithm returns to step (i) with the entire improved set of $\lambda[E_{\hbox{\scriptsize av}}]$.

{\it Algorithm A: Details and explanations}

At the small-$p_k$ approximation step, we substitute $\langle p_k \rangle$ from Eq.(\ref{pav2}) into Eq.(\ref{epsavav}), and then find numerically the value of $\lambda$ such that $E_{\hbox{\scriptsize av}} - {1 \over \lambda} < E_{\hbox{\scriptsize min}}$ or $E_{\hbox{\scriptsize av}} - {1 \over \lambda} > E_{\hbox{\scriptsize max}}$.

At the first iteration step, we start by substituting $\lambda[E_{\hbox{\scriptsize av}}]$ obtained in the small-$p_k$ approximation into the integral formula (\ref{Vkpk2}), thereby obtaining the generally non-exponential function $V_k(p_k)$, then use Eq.(\ref{pkav}) to obtain $\langle p_k \rangle$ and then check condition (\ref{epsavav}), which shows deviation from zero. In response to this deviation, we modify the function $\lambda[E_{\hbox{\scriptsize av}}]$. 

The value of $\lambda[E_{\hbox{\scriptsize av}}]$ at one grid point affects, according Eq.(\ref{Vkpk2}), the values of $V_k(p_k)$ and hence condition (\ref{epsavav}) at other grid points. Therefore, the proper iteration requires finding the whole matrix of linear responses of function
\begin{equation}
 F(E_{\hbox{\scriptsize av}}) = \sum_{k=1}^N  (E_k - E_{\hbox{\scriptsize av}}) \langle p_k \rangle
\label{F}
\end{equation}
at each grid point to the change of $\lambda$ at each other grid point. If the $E_{\hbox{\scriptsize av}}$-grid has $N_G$ points, then one has to solve a system of $N_G$ linear equations to find the modification $\Delta \lambda[E_{\hbox{\scriptsize av}}]$ for each grid point. From our experience such an approach often produces spurious solutions and, therefore, is difficult to manage. Instead, we adopt a different approach based on our understanding, why  the small-$p_k$ approximation works so well.

In the small-$p_k$ limit, the algorithm for finding $\lambda[E_{\hbox{\scriptsize av}}]$ is ``local'', i.e. it does not involve explicit influence of $\lambda[E_{\hbox{\scriptsize av}}]$ at a given grid point on $F(E_{\hbox{\scriptsize av}}^{\prime})$ at other grid points. In this case, the value of $\lambda$ is tried, and if it does not give $F(E_{\hbox{\scriptsize av}})=0$, then it is modified to $\lambda^{\prime} = \lambda + \Delta \lambda$, which, according to Eq.(\ref{Vkpk1}), implies that $V_k(p_k)$ is also modified to become
\begin{equation}
V_k^{\prime}(p_k) = V_k(0) e^{-N p_k[1 + (\lambda + \Delta \lambda) (E_k - E_{\hbox{\scriptsize av}})]}.
\label{VkpkP}
\end{equation}   
This equation can now be rewritten as
\begin{equation}
V_k^{\prime}(p_k) = V_k(p_k) e^{-N p_k\Delta \lambda (E_k - E_{\hbox{\scriptsize av}})]}.
\label{VkpkP1}
\end{equation}

In the above form, the iteration step is generalizable to the case, when $V_k(p_k)$ itself deviates from the exponential form. We find the value of $\Delta \lambda$ such that $F(E_{\hbox{\scriptsize av}})$ calculated with $\langle p_k \rangle$ corresponding to  $V_k^{\prime}(p_k)$ substituted into Eq.(\ref{pkav})  equals zero. Such an ansatz for  $V_k^{\prime}(p_k)$ is not fully consistent with Eq.(\ref{Vkpk2}), but it is a good approximation of it, and more importantly leads to a good improvement in the value of $\lambda$. The version of $V_k(p_k)$ consistent with Eq.(\ref{Vkpk2}) is then obtained at the beginning of the next iteration step  by substituting the entire improved function $\lambda[E_{\hbox{\scriptsize av}}]$ into Eq.(\ref{Vkpk2}), which, in turn, leads to the deviation of $F(E_{\hbox{\scriptsize av}})$ from zero, which is again canceled at each $E_{\hbox{\scriptsize av}}$ by replacing $V_k(p_k)$ by $V_k^{\prime}(p_k)$ with the new value $\Delta \lambda$, etc.
In our experience, the form of $\lambda[E_{\hbox{\scriptsize av}}]$ stops noticeably changing after three such iteration cycles. The calculated results presented in Section~\ref{results} are based on six iterations. Ultimately, it is the observed quick convergence that justifies the validity of the present algorithm.

The rationale for updating the values of $\lambda[E_{\hbox{\scriptsize av}}]$ on the basis of Eq.(\ref{VkpkP1}) is that (i) change of $\lambda$ at one value of $E_{\hbox{\scriptsize av}}$ affects mostly $F(E_{\hbox{\scriptsize av}}^{\prime})$ for $E_{\hbox{\scriptsize av}}^{\prime}$ reasonably close to $E_{\hbox{\scriptsize av}}$; and (ii) for $E_{\hbox{\scriptsize av}}^{\prime}$ close to $E_{\hbox{\scriptsize av}}$, the value of $\Delta \lambda[E_{\hbox{\scriptsize av}}^{\prime}]$ should not be very different from $\Delta \lambda[E_{\hbox{\scriptsize av}}]$ --- hence $\Delta \lambda[E_{\hbox{\scriptsize av}}]$ well approximates $\Delta \lambda[E_{\hbox{\scriptsize av}}^{\prime}]$ once the corrections are substituted in Eq.(\ref{Vkpk2}).
The correction $\Delta \lambda[E_{\hbox{\scriptsize av}}]$ obtained with the help of Eq.(\ref{VkpkP1}) accurately anticipates the linear term in the expansion of integral in Eq.(\ref{Vkpk2}) in powers of $p_k$ and the resulting exponential change of $V_k(p_k)$ for sufficiently small values of $p_k$. For the larger values of $p_k$, the exponential factor in Eq.(\ref{VkpkP1}) introduces a larger relative error, but, in almost all cases, this is a larger error to the small and hence less important tail of $V_k(p_k)$. The tail is then corrected at the beginning of the next iteration step once the updated function $\lambda[E_{\hbox{\scriptsize av}}]$ is substituted into Eq.(\ref{Vkpk2}).

{\it Algorithm B}

Algorithm A  can now be modified to incorporate the finite-$N$ corrections associated with the deviations of individual functions $\lambda_k[E_{\hbox{\scriptsize av}}]$ from the ``mean'' function $\lambda[E_{\hbox{\scriptsize av}}]$. 

Algorithm B continues searching for mean function $\lambda[E_{\hbox{\scriptsize av}}]$, but at each step of Algorithm~A that used Eq.(\ref{Vkpk2}), it uses Eq.(\ref{Vkpk2A}) with values of $\lambda_k[E_{\hbox{\scriptsize av}}]$ obtained from $\lambda[E_{\hbox{\scriptsize av}}]$ with the help of ansatz described in Section~\ref{ansatz}. The algorithm continues using the ``local'' step involving Eq.(\ref{VkpkP1}), and, here again, the fast convergence of the iteration procedure is the evidence of the adequacy of that step.

\section{Algorithm for direct Monte-Carlo sampling in energy-constrained Hilbert space}
\label{algorithmMC}

Without energy constraint (\ref{epsav}), there exist a straightforward and efficient algorithm for the sampling of Hilbert space. Namely, one can first choose all values of Re$(C_i)$ and Im$(C_i)$ according to the same Gaussian probability distribution. This gives unnormalized wave function. After it is normalized, the resulting Monte-Carlo points have the required isotropic probability density on the normalization manifold (provided one can control the effects of the cutoff on the tail of Gaussian distribution necessarily appearing in the numerical implementation of the sampling).

With energy constraint (\ref{epsav}), we are not aware of any algorithm, which would be as efficient as the above one and yet rigorously implement uniform sampling on the energy-normalization intersection manifold.   

We used the same algorithm as Ref.\cite{Fine-09-statistics}. Namely, we do the sampling in the $\{p_i\}$-Euclidean space, where we define an orthogonal basis in the $(N-2)$ dimensional subspace formed by the intersection of energy and normalization hyperplanes (\ref{epsav}) and (\ref{norm}).  The orientation of these hyperplanes are determined by the unit vectors, which are orthogonal to them:  $\mathbf{v}_E = {1 \over \sqrt{\sum_{i=1}^N E_i^2}} (E_1, E_2, ..., E_N)$ and  $\mathbf{v}_{\hbox{\scriptsize norm}} = {1 \over \sqrt{N}} (1,1, ...,1)$.  The basis within the intersection manifold is then generated by  Gramm-Schmidt orthogonalization procedure. Namely, we make the first basis vector $\mathbf{v}_1$ orthogonal to both $\mathbf{v}_E$ and $\mathbf{v}_{\hbox{\scriptsize norm}}$; then $\mathbf{v}_2$ orthogonal to $\mathbf{v}_E$, $\mathbf{v}_{\hbox{\scriptsize norm}}$ and $\mathbf{v}_1$; etc. Each of $(N-2)$ vectors $\mathbf{v}_i$ then gives rise to a new coordinate axis, which we denote as $q_i$. One has, of course, much freedom in choosing this basis. 

In the above $(N-2)$-dimensional subspace, the QMC polyhedron is further constrained by the positivity conditions $p_i \geq 0$. As a result, it has $KL$ vertices\cite{Fine-09-statistics}, where $K$, and $L$ are the numbers of energy levels in spectrum $\{ E_i \}$ above and below $E_{\hbox{\scriptsize av}}$, respectively. In the original basis of space $\{p_i\}$, each vertex has only two non-zero coordinates  
$(0, ..., 0,p_m,0,...,0,p_n,0,...,0)$ corresponding to one possible pair of energy levels such that $E_m < E_{\hbox{\scriptsize av}}$ and $E_n > E_{\hbox{\scriptsize av}}$. The values of these coordinates are $p_m = {E_n - E_{\hbox{\scriptsize av}} \over E_n - E_m}$ and $p_n = {E_m - E_{\hbox{\scriptsize av}} \over E_m - E_n}$. 

We choose the origin of the $q_i$-coordinate system on one of the above vertices.
The $p_i$ coordinates of each vertex are then transformed into $q_i$ coordinates.  Then, in the $q$-coordinate system, we choose an $(N-2)$-dimensional hyperrectangle (orthotope) with edges parallel to the $q_i$ axes and with the extension along each axis limited by $q_{i \hbox{\scriptsize min}}$ and $q_{i \hbox{\scriptsize max}}$ such that $q_i$-coordinate of each polyhedron vertex falls between $q_{i \hbox{\scriptsize min}}$ and $q_{i \hbox{\scriptsize max}}$, i.e. the polyhedron of interest is completely inside the above hyperrectangle.

The Monte-Carlo sampling is then straightforwardly implemented in the hyperrectangle by choosing each coordinate $q_i$  randomly in the interval $[q_{i \hbox{\scriptsize min}}, q_{i \hbox{\scriptsize max}} ]$.  The Monte-Carlo point is accepted, when, after transformation to the original $p_i$-coordinate system, all of its coordinates satisfy the positivity condition $p_i \geq 0$.

The acceptance rate of this algorithm decreases exponentially with increasing $N$, because high-dimensional polyhedron typically occupies exponentially small part of the volume of a hyperrectangle that covers it. 

The above ratio of volumes has a further strong dependence on the relative orientation of the hyperrectangle and the QMC polyhedron. It was our guess confirmed empirically that this ratio is strongly increased, once one of the faces of the polyhedron is parallel to one of the faces of the hyperrectangle.
The faces of the polyhedron are determined in the $p_i$-space by the intersection of three hyperplanes: $p_i =0$, energy hyperplane (\ref{epsav}) and normalization hyperplane (\ref{norm}). If one of vectors $\mathbf{v}_i$ is orthogonal to one of the polyhedron faces, then that face will be parallel to the corresponding face of the hyperrectangle. For example, if we want $\mathbf{v}_1$ to be orthogonal to the face determined by hyperplane $p_1 = 0$, then we choose $\mathbf{v}_1$ to be parallel to 
$\mathbf{u}_1 - (\mathbf{u}_1 \cdot \mathbf{v}_E) \mathbf{v}_E - (\mathbf{u}_1 \cdot \mathbf{v}_{\hbox{\scriptsize norm}}) \mathbf{v}_{\hbox{\scriptsize norm}} $,
where $\mathbf{u}_1 = (1,0,0,...,0)$ in the original $p_i$-space. For $N=12$, the resulting increase in algorithm's acceptance rate was, approximately, by a factor of 50.

We have not yet systematically optimized the choice of the ``parallel'' polyhedron face as a function of the corresponding energy $E_i$ from the viewpoint of increasing the polyhedron-to-hyperrectangle volume ratio, but, so far, our experience indicates that,  for $E_{\hbox{\scriptsize av}} < E_{\hbox{\scriptsize av}0}$, the face corresponding to the lowest energy, i.e. originating from hyperplane $p_1 =0$, is one of the best. The algorithm can, perhaps, be further improved by optimizing the subsequent choice of $\mathbf{v}_2$, $\mathbf{v}_3$, etc. from the viewpoint of increasing the above volume ratio.

\section*{References}


\end{document}